\documentclass[prd,groupedaddress,twocolumn,nofootinbib,showpacs,preprintnumbers,floatfix]{revtex4}

\usepackage{mathrsfs}
\usepackage{amsfonts,amssymb,amsmath}
\usepackage{graphicx}
\usepackage{multirow}
\usepackage{slashed}
\usepackage{fancybox}
\usepackage{subdepth}
\usepackage{float}
\usepackage{fancyvrb}
\usepackage{comment}
\usepackage[T1]{fontenc}
\usepackage{url}
\usepackage{hyperref}
\usepackage{breakurl}

\declareslashed{}{/}{0}{-0.05}{E}
\declareslashed{}{/}{0}{-0.05}{P}
\cornersize*{15pt}
\newfloat{card}{htb}{lop}
\floatname{card}{Card}
\makeatletter
\renewcommand\floatc@plain[2]{\setbox\@tempboxa\hbox{{\footnotesize {\bf {\@fs@cfont #1:}} #2}}%
\ifdim\wd\@tempboxa>\hsize {\@fs@cfont #1:} #2\par
\else\hbox to\hsize{\hfil\box\@tempboxa\hfil}\fi}
\@namedef{thecard}{\Alph{card}}
\makeatother

\newcommand{\dlhco}{\mbox{{\small {\itshape .\hspace{0.75pt}lhco }}}}

\begin{document}

\title{{\scshape CutLHCO}: A Consumer-Level Tool for Implementing\\Generic Collider Data Selection Cuts in the Search for New Physics}

\author{Joel W. Walker}

\affiliation{\mbox{Department of Physics, Sam Houston State University, Huntsville, TX 77341, USA}\\{\tt jwalker@shsu.edu} \hspace{9pt} {\tt www.joelwalker.net}}

%%%%%%%%%%%%%%%%%%%%%%%%%%%%%%%%%%%%%%%%%%%%%%%%%%%%%%%%%%%%%%%%%%%%%%%%%%%%

\begin{abstract}
A new computer program named {\sc CutLHCO} is introduced, whose function is the implementation
of generic data selection cuts on collider event specification files in the standardized \dlhco format.
This software is intended to fill an open market niche for a lightweight yet flexible ``consumer-level''
alternative to the {\sc Root} data analysis framework.  The primary envisioned application is as 
a filter on output produced by the {\sc PGS4} and {\sc Delphes} detector simulations, which
are themselves lightweight alternatives to the {\sc Geant4} based solutions favored by the large
LHC experiments.  All process control instructions are provided via a compact and powerful
card file input syntax that efficiently facilitates the reasonable approximation of most
event selection strategies and specialized discovery statistics commonly employed by the
CMS and ATLAS collaborations.  The structure, function, invocation and usage of the most recent
{\sc CutLHCO 2.0} program version are documented thoroughly, including a detailed deconstruction
of several example card file specifications.  The associated software is simultaneously
being made available for free public download.
\end{abstract}

%%%%%%%%%%%%%%%%%%%%%%%%%%%%%%%%%%%%%%%%%%%%%%%%%%%%%%%%%%%%%%%%%%%%%%%%%%%%


\maketitle

%%%%%%%%%%%%%%%%%%%%%%%%%%%%%%%%%%%%%%%%%%%%%%%%%%%%%%%%%%%%%%%%%%%%%%%%%%%%

\makeatletter
\close@column@grid
\onecolumngrid
\vspace{-8pt}
\begin{center}
\includegraphics[width=0.8\textwidth]{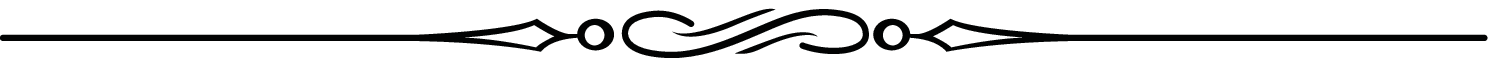}
\end{center}
\vspace{8pt}
\close@column@grid
\twocolumngrid
\makeatother

%%%%%%%%%%%%%%%%%%%%%%%%%%%%%%%%%%%%%%%%%%%%%%%%%%%%%%%%%%%%%%%%%%%%%%%%%%%%

\section{Introduction and Motivation}

%%%%%%%%%%%%%%%%%%%%%%%%%%%%%%%%%%%%%%%%%%%%%%%%%%%%%%%%%%%%%%%%%%%%%%%%%%%%

\subsection{Monte Carlo Event Generation}

In order to make a tangible connection between the abstraction of a given theoretical construct and the
actuality of detailed collider level observations, a sophisticated and reliable mechanism of simulation is essential.
However, the gulf separating the theoretical inception from the experimental inquest of a physical model can be quite wide.
The spanning of this great divide can be logically partitioned into four distinct steps, each of which may be sequentially
effected via widely established public computer code.

The initial computational task consists of generating low order Feynman diagrams that may link the incoming beam to the desired range of
hard scattering intermediate states.  The resulting matrix elements are subsequently fed into a secondary processing phase
for appropriate kinematic scaling and an implicit all-order resummation of the leading order transition into batches of Monte Carlo
simulated parton level scattering events.  In the case that the more precise but computationally demanding matrix element calculation
is extended to include radiation by higher order diagrams (as is increasingly advisable outside the soft, collinear secondary
emission regime), a one-to-one matching algorithm must also be applied to preclude the double counting of states.  Programs
popularly employed to facilitate this intricate sequence of calculations include {\sc AlpGen}~\cite{Mangano:2002ea}
and the {\sc MadGraph/MadEvent}~\cite{Stelzer:1994ta, Alwall:2007st} suite.

The third tier of processing, which handles the cascaded fragmentation and hadronization of the parton level events into final state
showers of photons, leptons and mixed jets, is the domain of programs such as {\sc Pythia}~\cite{Sjostrand:2006za},
{\sc Herwig}~\cite{Corcella:2000bw} and {\sc Sherpa}~\cite{Gleisberg:2008ta}.  Finally, though, a veil of obfuscation must be drawn
across the detailed omniscience of the initial three steps, replicating the limits on information and vulnerability to error of a physical
detector environment.  Options available for this fourth processing phase include the industrial strength {\sc Geant4}~\cite{Agostinelli:2002hh}
simulation toolkit favored by the major ATLAS and CMS detector collaborations at the Large Hadron Collider (LHC), and also lighter-weight
consumer-level tools such as {\sc PGS4}~\cite{PGS4} and {\sc Delphes}~\cite{Ovyn:2009tx}.

%%%%%%%%%%%%%%%%%%%%%%%%%%%%%%%%%%%%%%%%%%%%%%%%%%%%%%%%%%%%%%%%%%%%%%%%%%%%

\subsection{The Need for Data Selection Cuts}

The output produced by the utilities just described is not, however, directly suitable either for human consumption or for the efficient
discernment of signal from background.  For these purposes, a fifth processing phase is required -- one that is sensitive
to the extraordinary delicacy of the modern high energy experimental condition.  It is an environment in which known processes, sufficiently well
understood to be relegated to subservience as calibration, will, by their definition as the easier target, comprise a noise floor that tends to swamp
any candidate indication of new physics.  The severe synchrotron radiation limits on light particles has moreover forced circular ring collider probes at
the energy frontier to abandon the clean kinematic consumption of elemental electron-positron pairs for the muddled partial interactions of
strongly bound quark-gluon composites.  Any given set of final states, even assuming perfect efficiency in measurement, and admitting the
inevitable evanescence of the neutrino, will correspond to an innumerably large amalgam of unobservable internal processes.  The stochastic
variation inherent in quantum interactions will create false excesses and shortfalls in production which both mask and masquerade as the
sought post Standard Model (SM) contributions.  In this environment, discovery is only as certain as the aggregate of statistics, fastidiously
extracted from the tails of event distribution tails, accumulated with the laborious passage of time.

The most incisive tool in the particle physicist's arsenal for the clearing away
of these myriad obstacles is a savvy application of the data selection cut.
Simply put, it is necessary to isolate, or select, potential outcomes that are accessible to
the desired signal, but inaccessible to, or at least substantially unlikely for, the competing
background.  It may actually be beneficial to discard even a large quantity of signal
events from regions of phase space that are background dominated, in favor of a smaller
quantity of retained events of an unusual character that may be uniquely differentiated.
Additional layers of selection filtering may typically be applied to eliminate faked signals
attributable to the persistent incompleteness and occasional fallibility of detector
measurements.  Careful tailoring of the cuts to the sought signal may readily account for
orders of magnitude of relative signal enhancement, and great effort is thus expended in
this pursuit, commensurate with the weight of potential benefit that the cuts employed
may leverage against the great cost and effort of the project at large.

%%%%%%%%%%%%%%%%%%%%%%%%%%%%%%%%%%%%%%%%%%%%%%%%%%%%%%%%%%%%%%%%%%%%%%%%%%%%

\subsection{A New Selection Cut Tool}

Selection cuts are most often implemented by the major detector collaborations within the
{\sc Root}~\cite{Brun:1997pa} data analysis framework.  However, it would appear that there remains room 
within the Monte Carlo collider-detector simulation ecology for development of light-weight
consumer-level event selection tools that embody the analogous role with respect to {\sc Root}~\cite{Brun:1997pa} that 
{\sc PGS4}~\cite{PGS4} and {\sc Delphes}~\cite{Ovyn:2009tx} play with respect to {\sc Geant4}~\cite{Agostinelli:2002hh}.
Parallel efforts to fill this niche include the {\sc MadAnalysis5}~\cite{Conte:2012fm} package
available for integration with the {\sc MadGraph}~\cite{Alwall:2007st} software family, and {\sc Mathematica}
notebook based solutions such as the original {\sc Chameleon}~\cite{lhcowiki} package and various extensions
based upon it.  During the course of an extensive phenomenological study with colleagues Nanopoulos, Li and Maxin of a particle
physics model named $\cal{F}$-$SU(5)$~\cite{Maxin:2011hy, Li:2011fu, Li:2011av, Li:2012tr}, it proved beneficial 
to develop and refine a proprietary software solution for the implementation of desired cuts, and the counting
and compilation of the associated net statistics, in order to facilitate a clear and testable description
of the experimental profile that our preferred model might present at the LHC.  A documentation of the resulting
{\sc Perl} program, named {\sc CutLHCO} for its operation on event files in the standardized \dlhco output format that may
be produced natively by both {\sc PGS4}~\cite{PGS4} and {\sc Delphes}~\cite{Ovyn:2009tx}, is the object of the present article.

The {\sc CutLHCO} program has evolved organically during extensive private use over more than a year's time for the rapid
prototyping and application of generic selection cut criteria against Monte Carlo collider-detector event simulation data.
Development under the real-world pressure exerted by a need to carefully model a substantial (and growing) variety of actual 
selection strategies from the ATLAS and CMS collaborations has resulted in an extremely flexible and powerful analysis framework
that is capable of addressing the majority of leading contemporary use cases.  To cope with this extreme generality of function,
the user interface syntax has necessarily been driven toward an extremely simple and intuitive form, with all processing instructions
compactly delivered via a single input card file.  A recently established plateau in stability and sophistication has presented
an apt occasion for wider release of the associated package into the public domain under the terms of the GNU General Public
License~\cite{gnugpl}, in conjunction with assignment of the {\sc 2.0} versioning designation.  The full \mbox{{\sc CutLHCO 2.0}} distribution is
available for download from the author's personal website~\cite{cutlhco}, where it shall be updated as new versions become available.
The main program is also included as an ancillary file
\mbox{{\small \,``\!{\it anc}/\!{\it cut\_\!lhco.pl}\,''}} with this document's
electronic source at the arXiv.org repository.

%%%%%%%%%%%%%%%%%%%%%%%%%%%%%%%%%%%%%%%%%%%%%%%%%%%

\section{Program Structure and Function\label{sct:logic}}

Operation of the {\sc CutLHCO} program may be subdivided into three basic phases consisting
of the reading of an event specification in the \dlhco format (\ref{sct:lhco}), the filtering and
reconstruction of basic physics objects (\ref{sct:recon}), and the selection of event subsets
consistent with specific global signatures (\ref{sct:evtsel}).  An explicit deconstruction of the user
syntax required to control these various operational phases will be provided subsequently in Section~(\ref{sct:syntax}),
whereas the primary purpose of the present discussion is a higher-level introduction
to the logic, organization and physical underpinnings of the available program functionality.
As such, it may also serve more generically as a gentle contemporary review of the broad science
and art of the collider data selection cut.

%%%%%%%%%%%%%%%%%%%%%%%%%%%%%%%%%%%%%%%%%%%%%%%%%%%

\subsection{The \dlhco Object Format Input\label{sct:lhco}}

The processing cycle of a {\sc CutLHCO} program instance begins with the reading in of an \dlhco format event specification
file, which reasonably approximates the most critical aspects of object substructure that are knowable for a given collision within 
a modern high energy detector environment~\cite{lhcowiki}.  The physical geometry of such a particle detector generally includes 
an inner system of silicon pixel track finders with extraordinarily fine timing and position resolution surrounded by a network of
calorimeters designed to separately absorb energy through electromagnetic and hadronic interactions; intense magnetic fields curve
the paths of charged particles to facilitate a determination of the charge-to-mass ratio, and a supplementary outer gas detector may
be employed that specializes in the identification of muons, whose calorimeter depositions are generally weak.  An astute
integration of the raw data data collected by the various detector subsystems allows for the segregation of key particle species
with high confidence based upon various mutual tracking and calorimeter signatures, and also for the efficient handling of beam pile-up
conditions, multiple source interactions, intrusion from cosmic rays, and secondary vertex displacement (a key marker of heavy flavor physics). 

While still substantially less detailed than the corresponding {\sc Root}~\cite{Brun:1997pa} object classification, the \dlhco format
data subset represents a reasonable compromise between precision and ease of use.
The event constituents are combined into high-level physics objects consisting of photons, leptons (with specified flavor), and reconstructed
hadronic jet clusters.  These objects are described one per line in plain text, each with a variety of standard geometrical and kinematic
parameters provided in a reader-friendly fixed column layout.  All dimensionful quantities are given in GeV-scaled natural units.
Some degree of isolation filtering and cross-cleaning is typically applied prior to assembly of the high level objects.
It is likewise typical that decays in-flight of the tau have already been handled at a higher level of the analysis hierarchy,
and that tau leptons reported by the event specification are thus only the subset deduced from a hadronic signature within
the daughter product.  While it is usually possible to additionally restrict the \dlhco file content to events passing various
triggering criteria, it may be preferable to revert to only level-0 triggering in the detector simulation,
so that all event selection activity may be globally encapsulated as an exclusive domain of the {\sc CutLHCO} program.

In addition to the object type, each line of data includes an additional seven pieces of information, plus two dummy columns reserved
for future expansion of the format.  The first four serve as a one-to-one proxy for the energy-momentum 4-vector of the object,
translated into the traditional language of collider physicists. The pseudo-rapidity, defined as follows, 
is a pure function of the zenith angle $\theta$, as measured from the instantaneous $\hat{z}$ direction of travel of the
counterclockwise beam element.
\begin{equation}
\eta \equiv - \ln \tan(\theta/2)
\end{equation}
Forward (or backward) scattering correspond to $\eta$ equals plus (or minus) infinity,
while $\eta = 0$ is an entirely transverse scattering event.  The angle $\phi$ is simply the usual azimuthal angle, measuring
orientation around the beam axis.  The transverse momentum $P_{\rm T}$ gives the magnitude of the 3-vector momentum
$\vec{P}$ projection that is perpendicular to the beamline.
\begin{equation}
P_{\rm T} \equiv \sqrt{P_x^2+P_y^2}
\end{equation}
The final parcel of kinematic data is the object invariant mass $M$, which is especially important for jets representing the composition
of several lower level physical objects.  In particular, it should be recalled that the 4-vector sum of individually massless objects may
easily acquire an energy-momentum imbalance that manifests as a non-negligible mass-square in the invariant product.
\begin{equation}
P_\mu P^\mu \equiv E^2 - \vec{P}\cdot\vec{P} = M^2
\label{eq:minv}
\end{equation}
The given data may readily be inverted back into a standard 4-vector form for internal use.
\begin{eqnarray}
&\vec{P} \equiv \left\{ P_{\rm T} \cos \phi \,,\, P_{\rm T} \sin \phi \,,\, P_{\rm T} \div \tan \left( 2\, {\tan}^{-1} e^{-\eta} \right) \right\}&
\nonumber \\
&E \equiv \sqrt{\vec{P}\cdot\vec{P} + M^2}&
\nonumber \\
&P_\mu \equiv \left\{ E, \vec{P} \right\}&
\end{eqnarray}
It is also useful to define a quantity $\Delta R$ derived from the pairwise kinematic descriptions of two objects
that gives a fairly scaled measure of their relativistic ``angular separation'', in radians.
\begin{equation}
\Delta R \equiv \sqrt{ {( \Delta \eta)}^2 + {( \Delta \phi )}^2}
\label{eq:deltar}
\end{equation}

The final three active data columns contain additional information of a non-kinematic variety, and may perform multiple
duty by context.  One column is reserved to indicate the number of identified tracks associated with the object.  For
charged leptons, the sign of the track count distinguishes particles from anti-particles.  For hadronically reconstructed
taus, the track magnitude will be either 1 or 3.  For jets, it is a positive number or zero.  The next column is used
to indicate whether a given jet has been tagged for heavy quark flavor content (usually a $b$, but also sometimes $c$).
For muons, this column is appropriated instead to provide the numeric identity of the jet most adjacent in $\Delta R$
to the muon track.  It is often the case that this information will already have been employed by a muon cleaning
script at a higher level in the processing chain to remerge poorly isolated muons with a parent jet cluster.
If this has occurred, the appropriate jet kinematics will have been adjusted and the track count incremented for each
assimilated muon; additionally, the integer appearing after the decimal in the jet track count will reflect the number of
such occurrences.  The last data column provides the object's energy deposition ratio into the hadronic and electromagnetic
calorimeters.  For present purposes, it is useful to recast this ratio as an electromagnetic fraction ({\it cf.}~Ref.~\cite{PAS-SUS-09-001}).
\begin{equation}
\xi \equiv {\left(1+ E^{\rm had}/E^{\rm em}\right)}^{-1}
\label{eq:fem}
\end{equation}
Excessively large values ($\xi \sim 1$) may indicate that a given jet (especially one outside the pseudo-rapidity
bounds $\vert \eta \vert > 3$ where leptons and photons are typically explicitly classified) is actually an electron or a photon
rather than a hadronic object.  For muons, this column contains a decimal number, where the digits to the left of the decimal
indicate the amount of transverse momentum carried by adjacent tracks $P_{\rm T}^{\rm adj}$ within a $\Delta R \le 0.4$ cone around
the muon track, and the digits to the right of (and including) the decimal give a fraction representing the transverse calorimeter
energy $E_{\rm T}^{\rm cal}$ ({\it i.e.}~the energy deposition reduced by a trigonometric factor for angular orientation) 
within an inclusive three-by-three cell array surrounding the muon track divided by the muon transverse momentum.  Again, this
information is likely to have already been fed into a higher level processing phase dedicated to muon isolation cleaning, prior
to invocation of the {\sc CutLHCO} script.  For internal use, a composite transverse energy-momentum isolation ratio
({\it cf.}~Ref.~\cite{PAS-EWK-10-002}) is defined as follows. 
\begin{equation}
\zeta \equiv \left( P_{\rm T}^{\rm adj} + E_{\rm T}^{\rm cal} \right) / P_{\rm T}
\label{eq:temir}
\end{equation}

Subsequent to the itemization of constituent physics objects, the \dlhco format specifies a final line to report the composite
calorimeter based event missing transverse energy ${\slashed{E}} {}_{\rm T}^{\rm cal}$.  This is, again, primarily a calorimeter
based statistic, which is supplemented by additional kinematics extrapolated with aid of the muon detection subsystem.  In words,
this quantity measures the magnitude of the vector sum over directed calorimeter energy deposits, trigonometrically reduced to reflect
only components transverse to the beamline.  The imbalance reflected by a substantial ${\slashed{E}} {}_{\rm T}^{\rm cal}$ residue
indicates that weakly interacting particles may have escaped the detector.  In addition to the calorimeter estimate for the missing energy
vector magnitude, an angular orientation in the transverse plane is also provided; to be more precise, this orientation actually represents
the angular negation of the described vector sum, {\it i.e.}~the directionality of the missing energy deposition whose inclusion would
reclose the overall sum to the expected zero magnitude.  The topic of missing energy will be revisited in Subsection~(\ref{sct:evtsel}),
where the native \dlhco calorimeter based estimate will be contrasted to track based estimates computable directly within {\sc CutLHCO}. 

%%%%%%%%%%%%%%%%%%%%%%%%%%%%%%%%%%%%%%%%%%%%%%%%%%%

\subsection{Object Reconstruction\label{sct:recon}}

After reading the relevant content from the \dlhco input file, the {\sc CutLHCO} program
initiates an event processing phase that may be broadly labeled as ``object reconstruction'',
according to controls specified by the user in a card file.  The basic purpose of the object
reconstruction is to enforce minimum data quality characteristics for groups of leptons and jets,
in terms of their kinematics, geometry, proximity to other objects and multiplicity.  Classified
particle groups may be subsequently dereferenced by a numerical identifier during the program's
event selection phase, for testing against various discovery statistics designed to isolate new physics.
Elements of the card file syntax that are organically encountered during this dialog on the
object reconstruction procedure will be denoted by {\small {\tt Typewriter}} font for future reference.

The principal object grouping {\small {\tt OBJ\_ALL}} is, as the name suggests, applied universally during the
initial read access of all input physics objects.  It takes only two parameters, namely a global specification
of the acceptable boundaries on each object's transverse momentum magnitude $P_{\rm T}$ ({\small {\tt PTM}}) and
pseudo-rapidity magnitude $\vert \eta \vert$ ({\small {\tt PRM}}).  Objects failing this primary filtering are permanently
discarded from the event.  Next, a set of instructions {\small {\tt OBJ\_PHO}} is processed for the handling of photon
type objects.  In addition to the standard {\small {\tt PTM}} and {\small {\tt PRM}} filters, two additional parameters are made
available.  The {\small {\tt CUT}} parameter specifies the acceptable range of acceptable photon counts after application
of the previously described filters, as a minimum and maximum; noncompliance with the specified limits
causes the event to be rejected, or cut, as a whole.  The {\small {\tt JET}} parameter specifies how individual photon
candidates that fail the filtering specification are to be handled; a value of {\small ``{\tt 1}''} indicates that these
objects should be reclassified as jets, whereas the default behavior {\small ``{\tt 0}''} is simply to discard them.
The initial separation of the charged leptons proceeds in a very similar manner, as controlled by individual
object specifications {\small {\tt OBJ\_ELE}} ($e$), {\small {\tt OBJ\_MUO}} ($\mu$), and {\small {\tt OBJ\_TAU}} ($\tau$) for each generation.
The only universal operational distinction is the addition of an additional parameter {\small {\tt SGN}} for controlling the lepton charge;
values of {\small ``{\tt +1}''} and {\small ``{\tt -1}''} act to retain only the indicated sign grouping, whereas the default behavior
{\small ``{\tt 0}''} makes no distinction between particles and antiparticles.  For muons there are two additional specifications,
{\small {\tt PTC}} and {\small {\tt ETR}}, respectively describing the allowed range of transverse momentum $P_{\rm T}^{\rm adj}$
in GeV carried by tracks within a $\Delta R \le 0.4$ cone of adjacency to the muon, and the allowed range of dimensionless values
taken by the transverse energy-momentum isolation ratio $\zeta$ of Eq.~(\ref{eq:temir}).

From this point forward, the outer control structures refer to the leptons only as a unified class.  The
object reconstruction {\small {\tt OBJ\_LEP}} governs the content of the zeroth combined lepton classification, which constitutes
also the default starting point of all latter groupings.  Although the leptons are consolidated, finely
grained control over inclusion and exclusion by flavor is retained by the addition of a new parameter, {\small {\tt EMT}}.
This parameter can take the values {\small ``{\tt +1}''}, {\small ``{\tt +2}''} or {\small ``{\tt +3}''} to specify the inclusion only of
electrons, muons or taus, respectively, whereas the default behavior {\small ``{\tt 0}''} retains all lepton flavors.  However,
negative values {\small ``{\tt -1}''}, {\small ``{\tt -2}''} or {\small ``{\tt -3}''} are also allowed, and are interpreted instead as an exclusion
of the given flavor index; in practice, the value of {\small ``{\tt -3}''}, which selects both electrons and muons for retention,
is particularly useful.  The {\small {\tt SGN}}, {\small {\tt PTM}}, {\small {\tt PRM}} and {\small {\tt CUT}} parameters carry over in the expected manner,
but reclassification under the {\small {\tt JET}} flag is no longer permitted.  The {\small {\tt PTC}} and {\small {\tt ETR}} tags are also retained,
but affect only the muonic object subset.

The hadronic jets are treated very similarly to the unified lepton classifications, although there are distinctions in the available filtering
options.  The object reconstruction {\small {\tt OBJ\_JET}} governs the zeroth level jet assembly, which will again constitute the default
starting point for all subsequent groupings.  The {\small {\tt PTM}}, {\small {\tt PRM}} and {\small {\tt CUT}} parameters are supplemented by
four additional control specifications unique to the jets.  The parameter {\small {\tt HFT}} indicates whether jets are required to possess a heavy
flavor tag.  A value of {\small ``{\tt 1}''} indicates that loose tagging is sufficient, while a value of {\small ``{\tt 2}''} enforces
the stricter tight tagging criterion; the default setting {\small ``{\tt 0}''} enforces no requirements on heavy flavor content.
The parameter {\small {\tt FEM}} specifies a range of acceptable values for the dimensionless electromagnetic fraction $\xi$ of Eq.~(\ref{eq:fem}).
The tags {\small {\tt TRK}} and {\small {\tt MUO}} indicate the allowed range of counts, respectively, for individual object tracks associated
with the jet, and poorly isolated muons integrated into the jet by a prior cleaning phase.

The real power and flexibility of the {\tt CutLHCO} object reconstruction specification comes from the ability to
iteratively define a practically unlimited number of additional lepton and jet classification levels.  These go by
the control structure tags {\small {\tt OBJ\_LEP\_N}} and {\small {\tt OBJ\_JET\_N}}, where {\small {\tt N}} is positive integer between {\small ``{\tt 1}''}
and {\small ``{\tt 999}''}.  These specifications are processed in numerical sequence, and may refer for sourcing and comparison back
to any previously defined specification with a lower valued index.  Five new parameter keys are introduced,
in addition to the full set of retained {\small {\tt OBJ\_LEP}} and {\small {\tt OBJ\_JET}} inputs, to regulate this data network.
The most important is the source {\small {\tt SRC}}, which specifies the initial company of objects that are to
be filtered.  Indices provided for the lepton (jet) source parameter refer to previously defined lepton (jet) groups.
Multiple index values may be listed, and the union of all lepton or jet objects (as the case may be) included
within any of the referenced groups is assembled into the source for the current object reconstruction; negative
integers are also allowed, and have the effect of vetoing the inclusion of all objects present within any group so
referenced.  The default value {\small ``{\tt 0}''} adopts all objects within the zeroth lepton or jet classification
as the current source set.  A functionally similar parameter key {\small {\tt CMP}} is available for isolating a complementary
set of oppositely typed objects (leptons refer now to jets, and vice versa) against which a comparative filtering of objects 
may be applied.  Each of the {\small {\tt CMP}} and {\small {\tt SRC}} inputs is accompanied by a corresponding parameter
({\small {\tt CDR}} or {\small {\tt SDR}}) specifying the acceptable range of Eq.~(\ref{eq:deltar}) $\Delta R$ displacements between passed objects and
either the external complementary object set or the (otherwise successful elements of the) internal source object set.
Finally, the {\small {\tt ANY}} tag allows for multiple lepton or jet specifications to be joined by the logical construct ``{\bf or}''
rather than the default ``{\bf and}''.  The accepted input is a list of previously defined (lower valued) reconstruction
indices of the same lepton or jet variety as the host, with the effect that the linked set of selections is passed as a whole if
any single member (including the host) passes.  This is particularly useful for constructing the logical negation of a compound
statement, according to the identity
\mbox{${\rm \mathbf{not}}({\rm A}~{\rm \mathbf{and}}~{\rm B}) \equiv ({\rm \mathbf{not}}~{\rm A})~{\rm \mathbf{or}}~({\rm \mathbf{not}}~{\rm B})$}.
In particular, tiered classifications may be constructed that retain only events failing at least one of the criteria
simultaneously required for admission into a prior tier.

The final category of object reconstruction groupings {\small {\tt OBJ\_DIL\_N}} facilitates the identification of dilepton object pairs.
There is no default zeroth classification in this case, and the integer {\small {\tt N}} again ranges between {\small ``{\tt 1}''} and {\small ``{\tt 999}''}.
There are four new parameter specifications particular to the dilepton groupings, in addition to a {\small {\tt CUT}} on the number
of surviving pairs.  Firstly, {\small {\tt LEP}} specifies a single integer index corresponding to a previously defined source lepton
grouping.  Since the dileptons represent an independent data set, processed after all lepton classifications are complete,
the selected index does not have to be less than {\small {\tt N}}.  The zeroth lepton set {\small ``{\tt 0}''} is an acceptable specification,
but is not a default.  Next, {\small {\tt DLS}} specifies the target dilepton sign classification.  Allowed values are {\small ``{\tt +1}''}
for same-sign pairs only, {\small ``{\tt -1}''} for opposite-sign pairs only and the default value of {\small ``{\tt 0}''} to allow all combinations.
A similar specification {\small {\tt DLF}} restricts dilepton flavor mixing, where a value of {\small ``{\tt 1}''} allows only matching flavors
to be paired, and the default value of {\small ``{\tt 0}''} imposes no flavor restrictions.  Lastly, the specification {\small {\tt DMI}} provides
the acceptable range of the dilepton mass invariant $M^{\ell,\ell}$ in GeV, where the invariant mass $M^{\rm A,B}$ is defined
generically for any pair of objects A and B as follows.
\begin{eqnarray}
M^{\rm A,B} &\equiv& \sqrt{ \left( P^{\rm A}_\mu + P^{\rm B}_\mu \right) \left( P_{\rm A}^\mu + P_{\rm B}^\mu \right) }
\nonumber \\
&=& \sqrt{ M_{\rm A}^2 + M_{\rm B}^2 + 2\left( E^{\rm A} E^{\rm B} - {\vec{P}}^{\rm A} \cdot {\vec{P}}^{\rm B} \right)}
\nonumber \\
{\displaystyle \lim_{M_{\rm A} = M_{\rm B} = 0}} &\Rightarrow& \sqrt{ 2\, \vert {\vec{P}}^{\rm A}\vert \vert {\vec{P}}^{\rm B}\vert
\left( 1 - \cos {\Delta \varphi}^{\rm \,B,A} \right) }
\label{eq:minvab}
\end{eqnarray}
A massless limit for the individual leptons is generally a fair approximation in the present case, but is not enforced by the program.

%%%%%%%%%%%%%%%%%%%%%%%%%%%%%%%%%%%%%%%%%%%%%%%%%%%

\subsection{Event Selection\label{sct:evtsel}}

Following the reconstruction of component physics objects into groups satisfying various
data quality and counting requirements, the {\sc CutLHCO} program proceeds on with the
second half of its analysis duties, which may be summarized under the heading of ``event selection''.
This processing phase is dedicated to isolating global signatures of new physics that may become
apparent when the event is contemplated as a unified entity.  In particular, various discovery
statistics, either optimized for sensitivity to missing energy and exotic decay configurations or biased
against false triggers and energy mismeasurement, are evaluated under projection of the previously indexed
objects.  Elements of the card file syntax that are organically encountered as a part of this dialog on the
event selection procedure will be denoted by {\small {\tt Typewriter}} font for future reference.

%%%%%%%%%%%%%%%%%%%%%%%%%%%%%%%%%%%%%%%%%%%%%%%%%%%%%%%%%%%%%%%%%%%%%%%%%%%%

\subsubsection*{{\bf Selection Tier 1}}

Three elementary (but not fully independent) statistics are presently considered to comprise the first event selection tier,
each of which provides a distinct perspective on the basic transverse kinematics of the collision event.  The notion 
of transverse variables has been broached already in Subsection~(\ref{sct:lhco}), but their ubiquity of
application in the event selection process now comes front and center.  The driving motivation for adoption of this
perspective is the simple reality that the detector cannot be sensitive to particle flow down the beamline.  Moreover,
since the initial portion of longitudinal momentum carried by the interacting ``parton'' constituents in a hadron collision
cannot be ascertained, an imbalance in the detected momentum sum can only be established in the beam-transverse plane,
where the net conserved momentum may be safely approximated as zero.  This projection bestows an intrinsic kinematic
incompleteness on the transverse variables, by dint of which they will generically tend to indicate new physics via
a termination point rather than a peak.  Beyond this inherent loss of information, there are other obstacles that additionally
complicate the direct interpretation of transverse statistics.  As an important case in point, it is useful to consider the following
definition of the transverse energy of a single particle.
\begin{eqnarray}
E_{\rm T} &\equiv& \sqrt{ M^2 + {\vec{P}}_{\rm T} \cdot {\vec{P}}_{\rm T} } \,\,=\, \sqrt{E^2 - P_z^2}
\nonumber \\
{\displaystyle \lim_{M = 0}} &\Rightarrow& \vert {\vec{P}}_{\rm T}\vert
\label{eq:et}
\end{eqnarray}
Firstly, note that the characteristic Lorentz transformation response associated with energy as a time-like four-vector component
has been corrupted through a quadratic reduction (while retaining the same invariant mass $M$) by the longitudinal momentum.
However, this is not necessarily a detrimental exchange; in particular, the transverse energy $E_{\rm T}$
has acquired by trade the interesting new property of invariance under the subset of Lorentz transformations corresponding
to longitudinal boosts.  This is not a fundamental invariance of the type garnered by the mass $M$
in Eq.~(\ref{eq:minv}) from contraction under the Lorentz metric, but rather an ``accidental'' byproduct of the quiescence of
$\vec{P}_{\rm T}$ (which lacks time-like and longitudinal spatial components) under orthogonally directed space-time 
rotations; nevertheless, it is a very useful property that may be leveraged in interesting ways to link relative frames
of reference along the beamline.  Secondly, the mutual transverse energy $E_{\rm T}^{\rm A,B}$ of two summed four-vector
constituents A and B is not equivalent to the simple sum of individual transverse energies.
\begin{eqnarray}
{E}^{\rm A,B} &\equiv& {E}^{\rm A} + {E}^{\rm B}
\nonumber \\
\vec{P}{}^{\rm A,B} &\equiv& \vec{P}{}^{\rm A} + \vec{P}{}^{\rm B} \quad \& \quad \vec{P}{}_{\rm T}^{\rm A,B} \,\equiv\, \vec{P}{}_{\rm T}^{\rm A} + \vec{P}{}_{\rm T}^{\rm B}
\nonumber \\
E_{\rm T}^{\rm A,B} &\equiv& \sqrt{ (M^{\rm A, B})^2 + (\vert \vec{P}{}_{\rm T}^{\rm A,B} \vert )^2 } \,\ne\, E_{\rm T}^{\rm A} + E_{\rm T}^{\rm B}
\label{eq:epdiff}
\end{eqnarray}
In this sense, the transverse energy adds more like a mass than like an energy-momentum four-vector component.

\paragraph*{{\small {\bf $\boldsymbol{{\slashed{E}} {}_{\rm T}}$:}}}
The first, and arguably the most vital, of the dimensionful transverse statistics to be modeled for event analysis by the {\sc CutLHCO} program
is the missing transverse energy ${\slashed{E}} {}_{\rm T}$.  The inclusive calorimeter based estimate ${\slashed{E}} {}^{\rm cal}_{\rm T}$ of this quantity
described in Subsection~(\ref{sct:lhco}) is accessed through the selector tag {\small {\tt EVT\_CAL}}, and is accompanied by only a single 
control parameter, namely an optional {\small {\tt CUT}} in GeV.  However, in addition to this intrinsically available statistic, substantial
flexibility is provided to the user to craft transverse momentum $\vec{P}_{\rm T}$ track enhanced ${\slashed{E}} {}_{\rm T}$ variations as follows,
which may call upon preferred subsets of the event's physics objects.
\begin{eqnarray}
{\vec{{\slashed{P}}}} {}_{\rm T} &\equiv& -1 \times {\textstyle \sum_{i=1}^N} \left( P_{\rm T}^i \cos \phi_i \, \hat{x} + P_{\rm T}^i \sin \phi_i \, \hat{y} \right)
\nonumber \\
{\slashed{E}} {}_{\rm T} &\equiv& \vert {\vec{{\slashed{P}}}} {}_{\rm T} \vert
\,=\, \vert \vec{P}{}_{\rm T} \vert
\quad \left\{ {\displaystyle \lim_{\slashed{M} = 0 }} \right\}
\nonumber \\
{\slashed{E}} {}_{\rm T}^{\rm A,B} &\equiv& \left| {\vec{{\slashed{P}}}} {}_{\rm T}^{\rm A,B}
\equiv {\vec{{\slashed{P}}}} {}_{\rm T}^{\rm A} + {\vec{{\slashed{P}}}} {}_{\rm T}^{\rm B} \right|
\,\,\ne\,\, {\slashed{E}} {}_{\rm T}^{\rm A} + {\slashed{E}} {}_{\rm T}^{\rm B}
\label{eq:met}
\end{eqnarray}
Keeping with the previous verbal definition of ${\slashed{E}} {}^{\rm cal}_{\rm T}$, ${\slashed{E}} {}_{\rm T}$ is the magnitude of
the vector negation of the sum over the transverse momentum ${\vec{P}}_{\rm T}$ carried by $N$ individually reconstructed objects.
As before, it represents the unaccounted contribution required in order to balance 
the open polygonal vector sum of visible transverse momenta to the nominally expected net value of zero.
The composite missing transverse energy ${\slashed{E}} {}_{\rm T}^{\rm A,B}$ adds like the composite visible
transverse energy in Eq.~(\ref{eq:epdiff}) and is equal in magnitude to the mass-exclusive portion of that statistic, given that the missing
transverse momentum vector ${\vec{{\slashed{P}}}} {}_{\rm T}$ is the simple inverse of the visible transverse momentum vector $\vec{P}{}_{\rm T}$.
To recapitulate, the merit of this statistic lies in the fact that key indicators of new physics, particularly the SUSY neutralino $\tilde{\chi}$ dark matter
candidate, may interact only weakly and are expected to yield a missing energy signature upon escape from the detector.

It is in some circles conventional to distinguish estimates of this quantity that are augmented by data from the tracking subsystems as ``missing
transverse momentum'', while the term ``missing transverse energy'' is reserved for estimates based dominantly on directed calorimeter deposition;
one may even occasionally see a hybrid notation that superimposes a vector symbol onto the slashed energy.  Instead, given that there is no
unambiguous method for assigning a non-zero invariant mass $\slashed{M}$ to any underlying physics objects that may escape the detector network, the presently adopted
practice will be to designate all such vectorial space-like transverse imbalances as ${\vec{{\slashed{P}}}} {}_{\rm T}$, and all corresponding time-like
three-vector magnitudes as ${\slashed{E}} {}_{\rm T}$, resolving ambiguity as to the measurement origination by addition of the superscript ``cal'' when
applicable.  However, determination of a mass to be associated with the missing transverse energy remains a physically relevant issue, which
is driven by the practicality that escaping particles are frequently expected to be produced in pairs that may be entirely kinematically decoupled
from each other by intervening steps in the decay cascade, implying that a substantial portion of the unseen tracks associated with the true missing momentum
vector may internally cancel in the observable sum.  In fact, this dilemma forms the motivation for certain of the specialized discovery statistics to
be introduced shortly as the third event selection tier. 

The zeroth incarnation of the missing transverse energy ${\slashed{E}} {}_{\rm T}$ to be computed internally is labeled for input purposes as
{\small {\tt EVT\_MET}}, and includes the tracks of all constituents of the event that survive the primary {\small {\tt OBJ\_ALL}} classification.  It
is made available for reference to latter selectors by default, with no need for specific invocation by the user.  Again, an optional {\small {\tt CUT}}
may be specified in GeV.  Additional indexed {\small {\tt EVT\_MET\_N}} selectors, with {\small {\tt N}} an integer in the range {\small ``{\tt 1}''} to {\small ``{\tt 999}''},
are provided for evaluation of the missing transverse energy carried by various object subsets.  Two additional parameter
specifications, {\small {\tt LEP}} and {\small {\tt JET}}, allow sourcing from previously defined {\small {\tt OBJ\_LEP\_N}} and {\small {\tt OBJ\_JET\_N}}
object reconstructions. At least one of these inputs must be set equal to a single integer back-reference {\small {\tt N}}; a value of {\small ``{\tt 0}''} is allowed,
but is not a default.  The missing energy event statistics may themselves be made available for back-reference by their integer index to the {\small {\tt MET}}
parameter of subsequent selector definitions, with {\small {\tt EVT\_MET}} taking the implicit valuation of {\small {\tt N}~=~``{\tt 0}''}.  As a special
case, the calorimeter based estimate {\small {\tt EVT\_CAL}} is also automatically made available, under the index assignment {\small {\tt N}~=~``{\tt -1}''}.
The directional characteristics of each computed missing transverse energy statistic are retained internally for reference by future selectors.

\paragraph*{{\small {\bf $\boldsymbol{H_{\rm T}}$:}}}
A second event selector of great importance is the scalar sum on transverse energy $H_{\rm T}$, which is
defined as follows and indicated programmatically as {\small {\tt EVT\_MHT}}.
\begin{eqnarray}
H_{\rm T} &\equiv& {\textstyle \sum_{i=1}^N } E^{i}_{\rm T}
\nonumber \\
{\displaystyle \lim_{M_i = 0}} &\Rightarrow& {\textstyle \sum_{i=1}^N } \left| {\vec{P}}^{\,i}_{\rm T} \right|
\label{eq:mht}
\end{eqnarray}
A massless limit is imposed by default, but may be overridden by setting the {\small {\tt MAS}}
parameter to {\small ``{\tt 1}''}.  The key distinction between $H_{\rm T}$ and the missing transverse energy (in the massless limit)
is that the absolute value is applied prior to the summation rather than after, so that vector cancellation is avoided.  As such,
this statistic is instead representative of the net energy carried by the visible event constituents.  In fact, it represents
precisely the massive transverse longitudinal scalar possessing the property
of linear composition that eludes the transverse energy as defined by Eq.~(\ref{eq:epdiff}).
\begin{equation}
H_{\rm T}^{\rm A,B} \equiv  H_{\rm T}^{\rm A} + H_{\rm T}^{\rm B}
\label{eq:htdiff}
\end{equation}
In this sense, the massive limit might be considered somewhat more deeply consistent, insomuch as the desired associative property may be disrupted
by compound alternation of mass-zeroing with four-vector summation.  As before,
$H_{\rm T}$ is automatically computed for all primary event objects in the zeroth incarnation, and an optional
{\small {\tt CUT}} is allowed on the result in GeV.  Likewise, indexed {\small {\tt EVT\_MHT\_N}} $H_{\rm T}$ selectors
are available as before for treating {\small {\tt LEP}} and {\small {\tt JET}} object subsets.

\paragraph*{{\small {\bf $\boldsymbol{M_{\rm T}^{\rm eff}}$:}}}
The missing and visible energy magnitudes are combined in a third selector called the transverse effective mass $M_{\rm T}^{\rm eff}$,
which is defined as the following simple sum, and indicated programatically as {\small {\tt EVT\_MEF}}.
\begin{equation}
M_{\rm T}^{\rm eff} = {\slashed{E}} {}_{\rm T} + H_{\rm T}
\end{equation}
Once again, $M_{\rm T}^{\rm eff}$ is automatically computed for all
primary event objects, with a single optional parameter {\small {\tt CUT}}.  Additional indexed {\small {\tt EVT\_MEF\_N}} $M^{\rm eff}_{\rm T}$
selectors are likewise provided as before, but the supplemental parameter specifications are different in this
case.  Since $M_{\rm T}^{\rm eff}$ will be considered as a dependent construction based upon ${\slashed{E}} {}_{\rm T}$ and $H_{\rm T}$,
the available inputs are instead {\small {\tt MET}} and {\small {\tt MHT}}, each of which must specify an integer back-reference {\small {\tt N}} to
a previously defined incarnation of {\small {\tt EVT\_MET\_N}} or {\small {\tt EVT\_MHT\_N}}; a value of {\small ``{\tt 0}''} is allowed in
either case, but is not a default.

%%%%%%%%%%%%%%%%%%%%%%%%%%%%%%%%%%%%%%%%%%%%%%%%%%%%%%%%%%%%%%%%%%%%%%%%%%%%
 
\subsubsection*{{\bf Selection Tier 2}}

A second tier of derivative event selectors is constituted from various comparative combinations of the three
previously defined statistics.  No default analysis is performed in these cases, and an index {\small {\tt N}} in the usual
{\small ``{\tt 1}''} to {\small ``{\tt 999}''} range is always explicitly specified.  The first such statistic,
{\small {\tt EVT\_RET\_N}}, consists of the dimensionless comparison $( {\slashed{E}} {}_{\rm T}^{\rm N} / {\slashed{E}} {}_{\rm T}^{\rm D} )$
of two missing energy magnitudes.  This quantity might be used, for example,
to protect against the danger that reinclusion of discarded soft or high pseudo-rapidity jets could substantively alter 
the observed momentum imbalance.  Back-references to a pair of previously defined missing energy indices must be provided to
the inputs {\small {\tt NUM}} and {\small {\tt DEN}}, for use in the numerator and denominator, respectively.
The {\small {\tt CUT}} specification accepts pure numerical values for evaluation against the computed ratio.
A similar pair of selectors {\small {\tt EVT\_RHT\_N}} and {\small {\tt EVT\_REF\_N}} are provided for analysis of the
ratios $( {\slashed{E}} {}_{\rm T}^{\rm N} / \surd[ H_{\rm T}^{\rm D} ])$ and $( {\slashed{E}} {}_{\rm T}^{\rm N} / M_{\rm T}^{\rm eff\, D} )$,
which instead reference the scalar momentum sum and effective mass in the denominator, respectively.  The square root
employed in the {\small {\tt EVT\_RHT\_N}} example evinces the expected Gauss/Poisson fluctuation size, so that the statistic
provides a metric for the relative significance of the missing energy magnitude. 
The former {\small {\tt NUM}}, {\small {\tt DEN}} and {\small {\tt CUT}} inputs extend in a straightforward manner to each of these latter
two cases, although one should take note of the curious ${\rm GeV}^{1/2}$ units for the {\small {\tt EVT\_RHT\_N}} selector output.
The final statistic to be defined, {\small {\tt EVT\_DET\_N}}, represents the absolute vector difference
$\vert {\vec{{\slashed{P}}}} {}_{\rm T}^{\rm B} - {\vec{{\slashed{P}}}} {}_{\rm T}^{\rm A} \vert$
between two distinct population studies (A,B) of the missing transverse momentum.  Its function, once again, is to protect
against false indications of escaped particles.  In addition to a {\small {\tt CUT}} specification in GeV,
two parameters, {\small {\tt ONE}} and {\small {\tt TWO}}, are provided for specifying (in either order) a pair of
back-references to previously defined missing energy indices in the usual manner.

%%%%%%%%%%%%%%%%%%%%%%%%%%%%%%%%%%%%%%%%%%%%%%%%%%%%%%%%%%%%%%%%%%%%%%%%%%%%

\subsubsection*{{\bf Selection Tier 3}}

The third tier of provided event selectors represents a sampling of somewhat specialized statistics that are in contemporary 
use by the ATLAS and CMS collaborations.  As with the second tier, no default computation is performed in these cases, and an index
{\small {\tt N}} in the usual {\small ``{\tt 1}''} to {\small ``{\tt 999}''} range is always to be explicitly specified.  A robust variety of
analyses are provided with the presently documented {\sc CutLHCO 2.0} release, although the creation of additional subroutines to supplement the
generality of this coverage represents a potentially active source of ongoing program development.  The author welcomes useful suggestions for expansion
in this regard, and encourages users who write their own program extensions to resubmit them for inclusion into the main program trunk.

%%%%%%%%%%%%%%%%%%%%%%%%%%%%%%%%%%%%%%%%%%%%%%%%%%%

\paragraph*{{\small {\bf $\boldsymbol{M_{\rm T}^{\ell,{\slashed{E}}}}$:}}}
The first specialized statistic to be discussed is the transverse mass $M_{\rm T}^{\ell,{\slashed{E}}}$ computed for
a leading lepton object reconstruction and the missing transverse momentum, which shall be denoted for programmatic
purposes as {\small {\tt EVT\_LTM\_N}}.  This statistic has been employed frequently by the ATLAS
collaboration~\cite{ATLAS-CONF-2012-041,ATLAS-CONF-2011-130,ATLAS:2011ad}, although essentially similar channels are
also of interest to CMS~\cite{PAS-SUS-11-015}.
For any two objects A and B, the transverse mass statistic is defined as follows~\cite{kinematics}.
\begin{eqnarray}
M^{\rm A,B}_{\rm T} &\equiv& \sqrt{ (E^{\rm A}_{\rm T} + E^{\rm B}_{\rm T} )^2 - (\vert {\vec{P}}^{\rm A}_{\rm T} + {\vec{P}}^{\rm B}_{\rm T} \vert)^2}
\nonumber \\
&=& \sqrt{ M_{\rm A}^2 + M_{\rm B}^2 + 2\left( E^{\rm A}_{\rm T} E^{\rm B}_{\rm T} - {\vec{P}}^{\rm A}_{\rm T} \cdot {\vec{P}}^{\rm B}_{\rm T} \right)}
\nonumber \\
{\displaystyle \lim_{M_{\rm A} = M_{\rm B} = 0}} &\Rightarrow& \sqrt{ 2\, \vert {\vec{P}}^{\rm A}_{\rm T}\vert \vert {\vec{P}}^{\rm B}_{\rm T}\vert
\left( 1 - \cos {\Delta \phi}^{\rm \,B,A} \right) }
\label{eq:mt}
\end{eqnarray}
Note that the referenced transverse energy $E_{\rm T}$ of each constituent object is defined in a recursively consistent manner
according to the prescription by Eq.~(\ref{eq:et}).  Also, the simple sum of individual transverse energies that is featured in the
Eq.~(\ref{eq:mt}) definition mimics the $H_{\rm T}^{\rm A,B}$ prescription from Eq.~(\ref{eq:htdiff}) rather than the behavior of $E^{\rm A,B}_{\rm T}$ 
from Eq.~(\ref{eq:epdiff}).  In fact, by inspection, the mismatch between these quantities is precisely also the mismatch between the
pairwise invariant mass $M^{\rm A,B}$ of Eq.~(\ref{eq:minvab}) and its transverse variant $M_{\rm T}^{\rm A,B}$.

The $M_{\rm T}^{\ell,{\slashed{E}}}$ transverse mass distribution will have
an endpoint corresponding to the mass of the presumed common parent of the included lepton and missing energy
candidate, as may be verified by noting that \mbox{$M_{\rm T}^{\rm A,B} \le M^{\rm A,B}$}.  This potentially facilitates use as a tool
for the detailed mass measurement of SUSY particles, although the key present utilization is simply as a discovery statistic, where
a lower mass bound in the vicinity of 100~GeV is typically imposed.  The transverse mass $M^{\rm A,B}_{\rm T}$ is distinguished as a
Lorentz invariant under the subset of transformations consisting of longitudinal boosts for the same trivial reasons that the definition
of $E_{\rm T}$ from Eq.~(\ref{eq:et}) was itself so recognized.  An individually massless limit for both the lepton and missing transverse momentum
is enforced in the present $M_{\rm T}^{\ell,{\slashed{E}}}$ program implementation.
The available control parameters include the usual {\small {\tt CUT}} selector in GeV units, and index back-references
for the included {\small {\tt LEP}} object set (from which the leading lepton is extracted) and the counterpart {\small {\tt MET}} vector.

%%%%%%%%%%%%%%%%%%%%%%%%%%%%%%%%%%%%%%%%%%%%%%%%%%%

\paragraph*{{\small {\bf $\boldsymbol{M_{\rm T2}^{j,j}}$:}}}
A generalization of the previously described transverse mass statistic referred to as $M_{\rm T2}^{j,j}$ or the jet ``s-transverse mass'',
and which will be addressed programmatically as {\small {\tt EVT\_JSM\_N}},
has been adopted by the CMS collaboration~\cite{PAS-SUS-12-002} in order to better treat the SUSY scenario where pairwise sparticle production
leads to dual decay chains that each manifest a missing momentum signature.  First introduced in Ref.~\cite{Lester:1999tx},
the s-transverse mass construction is based upon the observation that each such visible shower may be convolved with its affiliated contribution to
the missing momentum to yield a transverse mass, as defined in Eq.~(\ref{eq:mt}), that is bounded above by the mass of the system's parent particle.
However, since only the unified missing transverse momentum vector ${\vec{{\slashed{P}}}} {}_{\rm T}$ is experimentally available,
further numerical analysis requires the specialization to a suitably optimized location in the distribution of this sum between the pair of
missing momentum candidates.  The $M_{\rm T2}$ prescription is to assume an equivalent structure for each decay chain, so that the larger of the two
transverse masses may be used imply a lower bound on a common parent species, subsequent to minimization of that maximum value with respect to all
consistent ${\vec{{\slashed{P}}}} {}_{\rm T}$ partitions.  If event pollution by upstream decays of the initial state prior to the targeted pair production
can be neglected, the minimization may actually be performed analytically~\cite{Cho:2007dh}.  If the mass of the escaping particle species $\tilde{\chi}$
is additionally set equal to zero, the following simple expression is realized.
\begin{eqnarray}
\Lambda_{\rm T} &\equiv& E^{\rm A}_{\rm T} E^{\rm B}_{\rm T} + {\vec{P}}^{\rm A}_{\rm T} \cdot {\vec{P}}^{\rm B}_{\rm T}
\nonumber \\
M^{\rm A,B}_{\rm T2} &\equiv& \sqrt{ \Lambda_{\rm T} + \sqrt{ ( \Lambda_{\rm T} )^2 - M_{\rm A}^2 M_{\rm B}^2 }}
\nonumber \\
{\displaystyle \lim_{M_{\rm A} = M_{\rm B} = 0}} &\Rightarrow& \sqrt{ 2\, \vert {\vec{P}}^{\rm A}_{\rm T}\vert \vert {\vec{P}}^{\rm B}_{\rm T}\vert
\left( 1 + \cos {\Delta \phi}^{\rm \,B,A} \right) }
\label{eq:mt2}
\end{eqnarray}
Relative to Eq.~(\ref{eq:mt}), one immediately notices the curious emergence of a Euclidean signature for the inner product.
This difference may be traced to the fact that $M_{\rm T2}$ employs two visible systems rather than a single visible system in
conjunction with associated missing energy; escaping missing energy is expected to be antiparallel to its conjugate visible shower,
inducing a phase shift by $\pi$ radians that effectively reverses the sign of the trigonometric factor. 
In order to apply the formula from Eq.~(\ref{eq:mt2}), the reconstructed objects taken to constitute the dual visible systems must be specified;
following the CMS lead~\cite{PAS-SUS-12-002}, {\sc CutLHCO} computes the s-transverse mass $M_{\rm T2}^{j,j}$ of a pair of optimally separated
pseudo-jets in the massless limit.  The ``event hemisphere reconstruction'' procedure ({\it cf.}~Ref.~\cite{Ball:2007zza}, Section~13.4)
begins by establishing a pair of pseudo-jet axis ``seeds'' consisting of the two individual massless jets that possess the largest relative
4-dimensional invariant mass.  These axes $P_\mu^{\rm A}$ are iteratively reconstructed as the 4-vector sum (with rezeroed mass)
of all massless jets $P_\mu^{\rm B}$ that are assigned to them in a given round by merit of the smaller Lund-inspired~\cite{Sjostrand:2006za}
dimensionful distance measure $(E_{\rm A} M_{\rm A,B}^2) / (E_{\rm A} + E_{\rm B})^2$.
The cycle terminates upon the retracing of a previously encountered hemisphere partitioning, which may (but does not necessarily) indicate
convergence to a stable fixed point assignment of the pseudo-jet axes.  Although originally introduced as tool for deducing the mass
scale of decaying SUSY particles, the s-transverse mass (like the transverse mass itself) is currently enjoying service as a discovery
statistic, with variously tuned cutoff thresholds ranging from around $100$~GeV up to about half a TeV~\cite{PAS-SUS-12-002}.
The available control parameters include only a {\small {\tt CUT}} selector in GeV units and an index back-reference
to the included {\small {\tt JET}} object set.

%%%%%%%%%%%%%%%%%%%%%%%%%%%%%%%%%%%%%%%%%%%%%%%%%%%

\paragraph*{{\small {\bf $\boldsymbol{\Delta E_{\rm T}^{j,\ell\ell}}$:}}}
The next specialized discovery statistic to be made available is the CMS jet and dilepton-Z transverse energy balance
$\Delta E_{\rm T}^{j,\ell\ell}$~\cite{Chatrchyan:2012qka}, which will be referenced by the tag {\small {\tt EVT\_JZB\_N}}.
Searches implementing this statistic are designed to target events featuring a leptonically decaying Z boson, which may be inferred by the
signature of residual opposite sign electron or muon pairs.  The production of Z bosons, in conjunction with jets and
missing energy, is a potentially regular byproduct of SUSY transitions from a heavy to light neutralino mass state.
Moreover, this selection naturally suppresses large portions of the SM event background, including the difficult QCD multijet component.
Troublesome opposite-sign dilepton background competition persists from the dual leptonic decay of pair produced top quarks
$t\bar{t}$, which may be obviated by leveraging the absence of flavor correlation in this process.  The $\Delta E_{\rm T}^{j,\ell\ell}$ statistic
constitutes an active discriminant against interference from SM Z bosons plus initial state jet radiation by highlighting the kinematic
correlation that may persist between the jet track momentum imbalance and the observed dilepton pair, if they indeed were borne of a
common antecedent.  It is formally defined as follows, employing a massless limit for each of the composite $N$-jet and dilepton objects.
\begin{eqnarray}
\Delta E_{\rm T}^{j,\ell\ell} &\equiv& \vert {\textstyle \sum_{j=1}^N}\, {\vec{P}}^{j}_{\rm T} \vert -
\vert {\textstyle \sum_{\ell = 1}^2}\, {\vec{P}}^{\ell}_{\rm T} \vert
\nonumber \\
&\approx& \vert {\vec{{\slashed{P}}}} {}_{\rm T} + {\vec{P}}^{\ell\ell}_{\rm T} \vert - \vert {\vec{P}}^{\ell\ell}_{\rm T} \vert
\label{eq:jdz}
\end{eqnarray}
The second line follows as a precise equality if the full missing momentum signal is attributable only to the dilepton and
included jets.  The unwanted background events are expected to be symmetrically balanced about the zero line of this statistic,
with random fluctuations driven in either direction by faked missing energy.  For the targeted SUSY events, directional
collinearity between the Z boson and the escaping light neutralino will systematically bias $\Delta E_{\rm T}^{j,\ell\ell}$ toward a
positive value.  The active signal region is typically restricted to that subset possessing a dilepton
invariant mass $M^{\ell,\ell}$ ({\it cf.}~Eq.~{\ref{eq:minvab}}) within about 20~GeV of $M_{\rm Z}$.  Additionally, one generally
requires the presence of at least three jets, to ensure decoupling between the orientation of potential momentum mismeasurements
and the Z boson track.  Along with a standard GeV {\small {\tt CUT}} selector, the user must specify back-reference indices for 
a pair of previously defined {\small {\tt JET}} and {\small {\tt DIL}} object sets.

%%%%%%%%%%%%%%%%%%%%%%%%%%%%%%%%%%%%%%%%%%%%%%%%%%%

\paragraph*{{\small {\bf $\boldsymbol{M_{\rm R}^{j,j}}$ \& $\boldsymbol{\alpha_{\rm R}^{j,j}}$:}}}
A trio of SUSY search statistics currently employed by the CMS collaboration~\cite{Chatrchyan:2011ek,PAS-SUS-12-005}
are the razor variables $M_{\rm TR}$ and $M_{\rm R}$, and their dimensionless squared ratio, denoted here as $\alpha_{\rm R}$.
Based upon a suggestion originally proffered in Ref.~\cite{Rogan:2010kb}, the pair of massive razor quantities
represent independent estimators, the former transverse and the latter longitudinally inclusive,
of the scale \mbox{$(m^2_{\tilde{q}}-m^2_{\tilde{\chi}})/m_{\tilde{q}}$} underlying compound SUSY event chains,
where $m_{\tilde{q}}$ is the common mass of an initial hard squark or gluino pair production event and $m_{\tilde{\chi}}$
represents a pair of weakly interacting particles from the dual decay cascade that escape the detector unseen.  In practice,
the jet razor mass variable $M_{\rm R}$, whose modeling is invoked by the tag {\small {\tt EVT\_JRM\_N}}, may be availed to 
isolate the scale of prospective new physics in conjunction with application of the $\alpha_{\rm R}$ ratio, which is provided
the independent identification tag {\small {\tt EVT\_ALR\_N}}, as a ``razor'' discriminant against SM background competition, primarily
from QCD di-jets.  The $M_{\rm TR}$ variable, which is dependent in this point of view, is not given its own accessor tag,
although the corresponding numerical value may be easily reconstructed as needed from the other two.
The razor statistics confront the same innate difficulty as the $M_{\rm T2}^{j,j}$ s-transverse mass analysis, namely that the
history of an event with a dual $\tilde{\chi}$ detection failure is essentially under-constrained by the residual visible particle content, generating
fundamental ambiguity within any reconstruction effort.  The strategy for surmounting this obstacle that is embodied by the razor
analysis rest upon an approximation of the opaque center of mass and pairwise $\tilde{q}$ rest frames (which will be identical in
the limit of pair-production at the kinematic threshold) via the expression of a transparent $\hat{z}$-axis boost into the ``razor frame'',
wherein the visible center of mass energy may be estimated in terms of the explicit longitudinal invariant $M_{\rm R}$ composed from
observations projected solely out of the laboratory frame;  the garnered longitudinal invariance protects the
integrity of the $M_{\rm R}$ distribution peak even if the boost into the true center of mass frame is non-trivial.
Construction of the razor frame is initiated by partitioning all relevant physics objects into two massless pseudo-jets ${\rm A}$ and ${\rm B}$
representing the observable remnant of the putative SUSY pair production event. The historically favored methodology 
is replicated by election of the object partition that minimizes the sum of pseudo-jet invariant mass-squares. The jets
are taken to be individually massless and, following their reassembly, the masses of each composite pseudo-jet
are likewise zeroed by an appropriate compensation of the energy.  The specific
required longitudinal boost is then guaranteed to exist and is expressible as \mbox{$\beta = (P^{\rm A}_z+P^{\rm B}_z)/(E^{\rm A}+E^{\rm B})$},
assuming vanishing of the transverse initial state radiation component.  The razor variables themselves are defined as follows.
\begin{eqnarray}
M^{\rm A,B}_{\rm TR} &\equiv& \sqrt{ \left\{ \vert{\vec{{\slashed{P}}}}{}_{\rm T} \vert\, (\vert {\vec{P}}^{\rm A}_{\rm T} \vert + \vert {\vec{P}}^{\rm B}_{\rm T} \vert) -
{\vec{{\slashed{P}}}}{}_{\rm T} \cdot ({\vec{P}}^{\rm A}_{\rm T} + {\vec{P}}^{\rm B}_{\rm T}) \right\} /\, 2 }
\nonumber \\
&=& \frac{1}{2} \sqrt{ ( M_{\rm T}^{\slashed{E},{\rm A}})^2 + ( M_{\rm T}^{\slashed{E},{\rm B}} )^2  }\quad \left\{ {\displaystyle \lim_{{\slashed{M}} = M_{\rm A} = M_{\rm B} = 0}} \right\}
\nonumber \\
M^{\rm A,B}_{\rm R} &\equiv& \sqrt{( E^{\rm A} + E^{\rm B} )^2 - ({P}^{\rm A}_{z} + {P}^{\rm B}_{z} )^2}
\nonumber \\
&=& \sqrt{ (M^{\rm A, B})^2 + (\vert \vec{P}{}_{\rm T}^{\rm A,B} \vert )^2 } \,\equiv\, E_{\rm T}^{\rm A,B}
\nonumber \\
\alpha_{\rm R} &\equiv& \left( M_{\rm TR} \,/\, M_{\rm R} \right)^2
\label{eq:razor}
\end{eqnarray}
In the transverse razor mass, it should be noted that a legitimate missing momentum, even if constructed inclusively from the event at large,
should be very nearly equal and opposite to the pseudo-jet vector sum, making the second term under the radical positive, and equal in magnitude
to the missing energy squared.  If the pseudo jets are close to parallel, then each term in the sum will moreover be equal, and $M_{\rm TR}^{j,j}$
will take the scale of the missing energy as a whole.  Conversely, if the pseudo-jets are back-to-back, as is increasingly likely
for QCD di-jets, a partial cancellation will be exhibited, even in the presence of faked missing energy.  If there is no missing energy signal at all,
then $M_{\rm TR}$ will be identically zero.  The first term under the $M_{\rm R}^{j,j}$ radical may be recognized as the Eq.~(\ref{eq:minvab})
invariant mass-square of the merged visible pseudo-jet system, and the expression as a whole is identical to the definition of the
transverse energy, as in Eqs.~(\ref{eq:et},\ref{eq:epdiff}), for this same system.  Since both mass estimators are designed to highlight the same
underlying physical scale, the ratio $\alpha_{\rm R}^{j,j}$ should be expected to peak at order one for signal events; specifically accounting
for ``geometric'' factors in the respective Eq.~(\ref{eq:razor}) definitions, the fact that $M_{\rm TR}$ (like all transverse statistics) identifies a kinematic endpoint,
and a small relative Lorentz transformation factor \mbox{$\gamma$~{\raise-.2ex\hbox{\small $\gtrsim$}}~$1$},
the signal peak will tend to occur closer to a value of $1/4$.
The SM QCD distribution will indeed peak at zero, allowing for a clear-cut excision of the encroaching background competition.
The available control parameters for each of the individually addressable statistics includes a {\small {\tt CUT}} selector in GeV units
and an index back-reference to the included {\small {\tt LEP}} and {\small {\tt JET}} object set;  for {\small {\tt EVT\_ALR\_N}} the user
may additionally include a reference {\small {\tt MET}} to the preferred missing transverse momentum, which will instead be rebuilt from
the local lepton and jet content if an index is not provided.

%%%%%%%%%%%%%%%%%%%%%%%%%%%%%%%%%%%%%%%%%%%%%%%%%%%

\paragraph*{{\small {\bf $\boldsymbol{\alpha_{\rm T}^{j,j}}$:}}}
A replication of the CMS $\alpha_{\rm T}$ ratio~\mbox{\cite{PAS-SUS-08-005,PAS-SUS-09-001,PAS-SUS-11-003}} is provided
for use under the identification handle {\small {\tt EVT\_ALT\_N}}.  This quantity was devised~\cite{Randall:2008rw} to help distinguish actual
missing transverse energy from detector mismeasurement. It is formally defined as follows,
where $\Delta H_{\rm T}^{\rm B,A}$ is the minimal positive difference that may be realized between the pair of $H_{\rm T}$ sums
corresponding to a grouping of all included objects into two effective pseudo-jets A and B, when considering all such possible partitions;
$H_{\rm T}^{\rm {\textstyle *}}$ is the lesser of $H_{\rm T}^{\rm A}$ and $H_{\rm T}^{\rm B}$.
\begin{eqnarray}
\alpha_{\rm T}^{\rm A,B} &\equiv& \frac{ H^{\rm {\textstyle *}}_{\rm T} }{ M_{\rm T}^{\rm A,B} }
\,=\, \frac{ \left( H_{\rm T}^{\rm A,B} - \Delta H_{\rm T}^{\rm B,A} \right) \!/\, 2 }{\sqrt{ (H_{\rm T}^{\rm A,B})^2 - ( {\slashed{E}} {}_{\rm T}^{\rm A,B} )^2 }}
\nonumber \\
&=& \frac{1}{2} \left\{ \frac{ 1- \left( \Delta H_{\rm T} / H_{\rm T} \right)}{\sqrt{ 1 - ( {\slashed{E}} {}_{\rm T} / H_{\rm T} )^2 }} \right\}
\label{eq:alphat}
\end{eqnarray}
If there is no mismeasurement or true missing energy, the value of $\alpha_{\rm T}^{j,j}$ will just be $1/2$.  For energy mismeasurements of
otherwise anti-parallel pseudo-jet pairs, subtraction of the nonvanishing scalar difference $\Delta H_{\rm T}^{j,j}$ will tend to drive $\alpha_{\rm T}^{j,j}$
below the midline.  In this case $\Delta H_{\rm T} = {\slashed{E}} {}_{\rm T}$, but the squaring of the small (typically {\raise+.2ex\hbox{\small $\ll$}}~$1$) factor in the denominator
renders it less significant.  Conversely, genuine missing energy, as manifest in the departure from pseudo-jet anti-parallelism, will imbalance the vector sum
within ${\slashed{E}} {}_{\rm T}$ more so than the simple magnitude difference $\Delta H_{\rm T}$, tending to create a contrasting elevation in
$\alpha_{\rm T}^{j,j}$ above one-half.  In addition to a dimensionless {\small {\tt CUT}} selector, the user may specify index back-references
for the included {\small {\tt JET}} object set, and the referenced {\small {\tt MET}} and {\small {\tt MHT}} values.  The latter two quantities will be
recomputed from the jet content itself if omitted.  Additionally, the same {\small {\tt MAS}} tag available to {\small {\tt MHT}} itself is also available
here to indicate a preference for the mass inclusive determination of $H_{\rm T}$.  As an editorial aside, it has been observed by the author's research
collaboration~\cite{Maxin:2011hy,Li:2011fu} that the $\alpha_{\rm T}$ statistic may become
unsuitable in the soft, high multiplicity jet regime, as the large array of finely grained
pseudo-jet permutations makes quite likely that a balanced scalar sum $\Delta H_{\rm T} \simeq 0$ might be achieved,
creating an overall suppression of the final computed value.

%%%%%%%%%%%%%%%%%%%%%%%%%%%%%%%%%%%%%%%%%%%%%%%%%%%

\paragraph*{{\small {\bf $\boldsymbol{\Delta \phi{}_{j,\slashed{E}}}$:}}}
The next event selector to be discussed is a relatively simple angular difference statistic $\Delta\phi{}_{j,\slashed{E}}$ employed by
the ATLAS~\cite{ATLAS-CONF-2012-033} collaboration.  It will be labeled {\small {\tt EVT\_MDP\_N}}, and is defined as follows.
\begin{equation}
\Delta \phi{}_{j,\slashed{E}} \equiv {\textstyle \min {}_{j=1}^N} \left\{ \left\lvert\,
\phi \left( {\vec{{\slashed{P}}}} {}_{\rm T} \right) - \phi\left( \vec{P}{}_{\rm T}^{\,j} \right) \right\rvert \right\}
\end{equation}
To summarize, $\Delta\phi{}_{j,\slashed{E}}$ simply isolates the minimal angular separation in the transverse plane between a
set of $N$ reconstructed objects and a fixed missing energy vector.  It is designed to protect against spurious indications of
missing energy associated with detector mismeasurement.  The user may specify a {\small {\tt CUT}} on
the resulting $0$ to $\pi$ radian angle, as well as index back-references for the active {\small {\tt JET}} object set and the
static {\small {\tt MET}} vector, the latter of which will be reconstructed from the included jets if omitted.
To reduce the likelihood of randomly generating a dispositively small value for $\Delta\phi{}_{j,\slashed{E}}$ concurrently 
with a legitimate missing energy signal, this statistic is typically computed for no more than the four hardest jets.

%%%%%%%%%%%%%%%%%%%%%%%%%%%%%%%%%%%%%%%%%%%%%%%%%%%

\paragraph*{{\small {\bf $\boldsymbol{\Delta \phi{}^{\rm {\textstyle *}}_{j,\slashed{E}}}$:}}}
The final event selector to be modeled is a CMS statistic~\cite{PAS-SUS-09-001}
that is quite similar to the just discussed angular difference $\Delta\phi{}_{j,\slashed{E}}$,
but with a somewhat more complicated implementation.
This ``biased'' azimuthal difference $\Delta \phi{}^{\rm {\textstyle *}}_{j,\slashed{E}}$, is formally defined as follows, and
will be associated with the program invocation tag {\small {\tt EVT\_BDP\_N}}.
\begin{equation}
\Delta \phi{}^{\rm {\textstyle *}}_{j,\slashed{E}} \equiv {\textstyle \min {}_{j=1}^N} \left\{ \left\lvert\,
\phi \left( {\vec{{\slashed{P}}}}{}_{\rm T} + \vec{P} {}_{\rm T}^{\,j} \right) - \phi\left( \vec{P}{}_{\rm T}^{\,j} \right) \right\rvert \right\}
\end{equation}
In words, $\Delta\phi{}^{\rm {\textstyle *}}_{j,\slashed{E}}$ is the minimal value of the set of $N$ angular separations in the transverse plane between each
reconstructed object's own momentum vector and the missing transverse momentum composed from all other objects.  It is
expressed as an angle in radians, in the range $0$ to $\pi$.  Intuitively, it registers whether the elongation of a single
momentum vector is capable of substantially repairing an imbalance in the missing transverse momentum.  If a single jet mismeasurement
is indeed dominantly responsible for a false missing energy signal, then $\Delta \phi{}^{\rm {\textstyle *}}_{j,\slashed{E}}$ should register close to zero.
In addition to an angular {\small {\tt CUT}} selector in radians, the user may specify index back-references
for the included {\small {\tt JET}} object set, and the baseline {\small {\tt MET}} vector.  If the latter quantity is omitted,
it will be recomputed from the jet content itself.  This statistic may again suffer in the case of very high jet multiplicities;
given a wide selection of randomly oriented jets, it becomes quite likely that the angular orientation of at least one such member might be sufficiently well
azimuthally aligned with the true missing energy track that its rescaling could apparently rebalance the event.

%%%%%%%%%%%%%%%%%%%%%%%%%%%%%%%%%%%%%%%%%%%%%%%%%%%

\section{Program Invocation and Usage\label{sct:syntax}}

%%%%%%%%%%%%%%%%%%%%%%%%%%%%%%%%%%%%%%%%%%%%%%%%%%%

\subsection{Card File Command Syntax}

All object reconstruction and event selection operations available to the {\sc CutLHCO}
program are controlled by command lines specified in an external card file.  An overview
of the logic and scope of the available functionality was provided previously, in
Section~(\ref{sct:logic}).  Five exemplar cards from the dozens of working selection
strategies accompanying the current program distribution~\cite{cutlhco} are documented
in the forthcoming Section~(\ref{sct:cards}), with a line-by-line textual deconstruction
of the specific command formatting and intent provided in each case.  A compact yet
comprehensive synopsis of the command lexicon is provided in the Appendix.  The
purpose of the present section is a description of the uniform aspects of the
card file command syntax and grammar.

Each processing instruction is placed on its own line in the card file
and begins by invoking the name of an object reconstruction ({\small {\tt OBJ\_ABC\_N}})
or event selection ({\small {\tt EVT\_ABC\_N}}) identifier, where {\small {\tt ABC}}
represents a three character alphabetic code unique to each command family, and 
{\small {\tt N}} is an integer within the range {\small ``{\tt 0}''} to {\small ``{\tt 999}''}.
Omission of the integer {\small {\tt N}} and the associated trailing underscore is functionally
equivalent to a selection of {\small {\tt 0}}.  This zeroth integer specification is generally
reserved for quantities that are automatically evaluated by the program, irrespective of user input,
although it may still be possible to modify the default parameter configuration in these cases through
an explicit reference.  However, not all input tags provide a meaningful zeroth configuration, and, conversely,
not all of those tag specifications that do automatically provide for a zeroth configuration additionally
support multiple numerical versioning.  Any lines of input that do not begin in the manner described
are disregarded as comments.  Likewise, the inclusion of a {\small ``{\tt \#}''} hash symbol
at the end of a valid input line is interpreted to initiate a comment enveloping all
trailing content.  The order in which lines are specified in the card file is not material
to the sequence in which they are ultimately evaluated;  the order of program execution is
instead established by the fixed operations sequence described in Section~(\ref{sct:logic}),
and within that framework, whenever there might otherwise be ambiguities, by an ascending numerical
sort on the associated numerical tags.

The parameter specification input for each command line is separated from the leading identifier name by
an equals sign {\small ``{\tt =}''}.  This parameter specification consists of a list of key and value
pairs, individually joined by a colon {\small ``{\tt :}''}, and separated from adjacent pairs by a
comma {\small ``{\tt ,}''}.  Each parameter key consists of a three character alphabetic string
that uniquely specifies the role of the subsequently input value.  The valid parameter specifications
vary according to the specific object reconstruction or event selector context of each command identifier,
as detailed in the document Appendix.  The order in which a given command line's parameter specifications are
provided is a matter of the user's discretion and has no impact on the logical flow of the selection algorithms.
The value assigned to each key may consist of an integer or floating point number (including signed values),
the character string {\small ``{\tt UNDEF}''} (to specify an undefined or null input), or a list consisting of
elements drawn from the prior classes, individually separated by commas {\small ``{\tt ,}''}, and enclosed in
square brackets {\small ``{\tt [~\ldots~]}''}.  The set of value inputs that may be considered valid or useful is,
again, a function of the context provided by the active parameter key.  Unspecified values default to
{\small ``{\tt UNDEF}''}, or if the expected input is an integer to be chosen from a predefined list of
options, the value {\small ``{\tt 0}''}.  Parameters designed to accept multiple inputs will treat an isolated
value as the leading list element, with all other terms undefined; likewise, there is no penalty for enclosing
a single expected input in square brackets, creating a list composed of a single entry.

A frequently employed value formatting idiom is the {\small ``{\tt [Min,Max]}''} pair, used to indicate
a numeric range.  So long as {\small {\tt Min}}{\footnotesize ~$\le$~}{\small {\tt Max}}, the set of matched numbers simply
consists of those values inclusively bounded by the specified range.  An undefined value for {\small {\tt Min}}
is treated as indefinitely small, and an undefined value for {\small {\tt Max}} is treated as indefinitely
large.  If no numerically valid limits are provided, then all values match.  If the upper and lower boundaries
are numerically equivalent, then only that single value matches.  However, a subtle additional functionality
is accessed under the circumstance that {\small {\tt Min}}{\footnotesize ~$>$~}{\small {\tt Max}}.  In this case, a numerical
match is achieved if the comparison value is either at least as large as {\small {\tt Min}} or at least
as small as {\small {\tt Max}}; in other words, the interval from {\small {\tt Max}} up to {\small {\tt Min}}
is rejected, exclusive of the boundaries.  Often, an additional instruction flag is accepted from the third position of
the described value list, which may be used to indicate various processing preferences, according to context.
The availability and function of this third input, whose uses include sorting objects that match 
the specified range criteria, terminating object matching upon an initial failure, and clipping the
count of objects satisfying some independent selection, is efficiently summarized as part of the encompassing
command specification reference provided as a document Appendix.

%%%%%%%%%%%%%%%%%%%%%%%%%%%%%%%%%%%%%%%%%%%%%%%%%%%

\subsection{Running the Program}

The {\sc CutLHCO} ``executable'' is delivered as a single {\sc Perl} script named {\small {\it cut\_lhco.pl}\,}; to be
more precise, since {\sc Perl} is an ``interpreted'' language, this file serves both as the program source
document and runtime portal.  A benefit of this paradigm is inherent platform independence, making the
{\sc CutLHCO} program ready for immediate use, without the need for installation or compilation, on any computer 
with a reasonably modern {\small {\it perl}} environment; this should automatically include almost all systems in current
service that run a Unix/Linux flavor variant, including Mac~OS~X, and free installations for Windows are also
readily available on the web.  The detailed instructions to follow will assume a Unix styled directory structure
and command line shell.

The {\sc CutLHCO} distribution package is available for download from the web~\cite{cutlhco}, and is delivered as a tar-gzipped
file bundle.  To unpack the distribution, the user should run the commands {\small ``{\tt gunzip cut\_lhco.tar.gz}''} and
{\small ``{\tt tar -xf cut\_lhco.tar}''} in sequence from the command line.  The directory from which {\sc CutLHCO} is called
must possess two specific subdirectories named {\small {\it Events}/} and
{\small {\it Cards}/}, which will house, respectively, the \dlhco format input files and the
card files controlling the application of selection cuts.  These directories are created automatically upon unpacking the
program distribution as described.  Users of the {\sc MadGraph}~\cite{Alwall:2007st}
simulation environment will recognize this directory structure, which is in fact appropriated to streamline integration with that
framework.  At the user's discretion, the main program file may likewise be grouped with any existing scripts and executables
in a common {\small {\it bin}/} subdirectory.  Execution of the script can then be as simple as entering {\small ``{\tt ./cut\_lhco.pl}''}, or
{\small ``{\tt ./bin/cut\_lhco.pl}''}, at the terminal window, as the case may be; however, there are several optional command-line parameters that
the user may wish to subsequently specify, which shall now be described in the sequence of their usage.  Any such instructions
are typed directly after invocation of the program name, separated from it and from each other by an empty space;  any parameter
position may be effectively skipped by providing the value {\small ``{\tt UNDEF}''}.

The initial parameter input is associated with the {\small {\it filename}} that one wishes to process, and it takes a default value
of {\small ``{\it data}\,''}.  Any processed output will ultimately be placed into a corresponding file named
\mbox{{\small ``{\it Events}/\!{\it filename.cut}\,''}}.  The referenced input file must physically exist within the {\small {\it Events}/} directory, either as
\mbox{{\small ``{\it filename.lhco}\,''}} or as \mbox{{\small ``{\it filename\_pgs\_events.lhco}\,''}}.  However, the latter scenario actually allows for two 
possible points of flexibility.  A group of files within {\small {\it Events}/} that follows the specified root name with an extra
underscore and a positive integer, as may be produced by the {\sc MadGraph}~\cite{Alwall:2007st} helper script {\small ``{\it multi\_run}\,''},
will be treated as a family of batch jobs to be merged for analysis.  Also, any matching files with a trailing extension of  
{\small \mbox{{\it .\hspace{0.75pt}gz}}} will be automatically unpacked with the {\small {\it gunzip}} utility prior to subsequent processing.
Additionally, the more simply named {\small ``{\it filename.lhco}\,''} file will be created, and filled with the appropriate merged
content.  The next parameter input position is associated with the {\small {\it cardfile}} governing event selection, and has
a default value of {\small ``{\it cut\_card}\,''}.  The referenced file must likewise exist within the {\small {\it Cards}/} directory, as
{\small ``{\it cardfile.dat}\,''}.

The third parameter position allows the user to specify a numerical cross section in
${\rm pb}$ for the physical processes under consideration.  Specifying the negative cross-section value {\small ``{\it -1}\,''} causes
the program to search for a file named {\small ``{\it filename\_pythia.log}\,''} (or the set of integrally labeled file variants)
in the {\small {\it Events}/} directory, from which the cross section will be dynamically extracted, along with a statistic
for the fragmentation cut percentage.  An unrecoverable error signal is issued if the expected file (or files) are not found.
If multiple records are queried, the cross section and fragmentation cut values are averaged, with equal weight assigned to
each individual file.  The fourth position designates a target luminosity
in ${\rm pb}^{-1}$ to which the surviving event count should be scaled.  Since this calculation is dependent upon
a known cross section, the reading of this input is skipped unless the third parameter was defined and non-zero; if
required but not defined, the luminosity target is set to a default of $1,000~{\rm pb}^{-1} \equiv 1.0~{\rm fb}^{-1}$.
The fifth and final input, which may actually shift into the fourth position as described, represents an integer variable
for controlling the output of detailed event-by-event statistics.  The default value {\small ``{\tt 0}''} suppresses this
output, while {\small ``{\tt 1}''} enables it, and {\small ``{\tt 2}''} additionally implements a resorting of the original event sequence
by the numerical value of each data column, from left to right; this sorting process may take an appreciable amount of time for large
quantities of surviving events with large multiplicities of statistics.

To provide a concrete initiation example, the command {\small ``{\tt ./cut\_lhco.pl ttbar cut\_card\_LHC 79.8 4500 1}''} will,
given placement of the main executable within the working directory, cause
a file named {\small ``{\it ttbar.lhco}\,''} (or its permissible variants) within the {\small {\it Events}/} directory to be processed according to the
selection cut strategy described in the file {\small ``{\it Cards}/\!{\it cut\_\!card\_\!LHC.dat}\,''}, assuming a process cross-section of
$79.8~{\rm pb}$, with results scaled up to a sample size of $4.5~{\rm fb}^{-1}$, and all event statistics reported but not sorted.

%%%%%%%%%%%%%%%%%%%%%%%%%%%%%%%%%%%%%%%%%%%%%%%%%%%

\subsection{Program Output and Interpretation}

The output {\small \mbox{{\it .\hspace{0.75pt}cut }}} file described in the prior section provides a summary report
of the {\sc CutLHCO} runtime environment and any actionable statistics generated from the surviving events.
It opens by stating the number of \dlhco files merged during the process initiation (if
there were multiple inputs), the total number $N$ of events processed, the percentage of events failing
fragmentation cuts (if harvested from the {\sc Pythia}~\cite{Sjostrand:2006za} log),
the source event cross section $\sigma$ in ${\rm pb}$ and associated simulation luminosity ${\cal L} = N/\sigma$
in ${\rm pb}^{-1}$ (if available), and the target luminosity and associated event rescaling (if non-trivial). 
These particulars of the operational context are followed by a presentation of key results and analysis, beginning with
the all-important scaled count of events surviving the selection cut regimen.  Next, the percentage of events cut
during the selection is tabulated, as a unified total and also as a statement of activity for each individually
defined cut; the latter set of statistics is supplemented by the percentage tally of circumstances in which each 
specified cut constitutes a uniquely enforced rationale for event exclusion.  Each object or event command
providing a value for the {\small ``{\tt CUT}''} selector that is not functionally equivalent to
{\small ``{\tt UNDEF}''} will be included in this report, referenced by its unique alphabetic identification code
(omitting the {\small ``{\tt OBJ\_}''} or {\small ``{\tt EVT\_}''} prefix) and its numerical sequencing tag.

If a detailed itemization of results was specified during the program initiation, then each surviving
event will be given a line in the output that inventories its response to each defined selection cut query.
This is particularly useful for tabulating histogram reports in various discovery statistics, or for partitioning
the counts of events satisfying subordinate selection conditions.  It may also be usefully applied to optimization
of the underlying selection strategy itself, facilitating the identification of inflections or knees in the event
distribution, potential advantages over (or novel tools for the suppression of) background, and selection redundancies
or inefficiencies.  To include a given output statistic in this report without actually imposing a cut, a trivial value
should be provided for the {\small {\tt CUT}} selector that is still technically defined, {\it e.g.}~an inclusive
lower bound of {\small ``{\tt 0}''} for an inherently positive quantity such as $P_{\rm T}$.  A helper script
{\small {\it histogram.pl}} is provided in the main directory of the {\sc CutLHCO} distribution~\cite{cutlhco} that is well
suited for this analysis task.  This script concurrently processes all {\small \mbox{{\it .\hspace{0.75pt}cut }}} format files within
the {\small {\it Events}/} subdirectory, and place the results into a file called {\small ``{\it histogram.txt}\,''} in the working directory.

The histogramming script accepts up to six optional space-separated parameter inputs after the primary invocation
{\small ``{\tt ./histogram.pl}''}, and again permits the usage of {\small ``{\tt UNDEF}''} for reversion to system defaults.
The first input should be the name of the event detail column heading from the relevant {\small \mbox{{\it .\hspace{0.75pt}cut }}}
files that the histogram should be based upon.  The default behavior is a simple single-column histogram on the net
event count; although this repeats raw information already provided in the underlying report, it may still be useful
for rapidly assembling the counts of multiple event samples into a unified presentation.  The second input is a numerical
specification for the binning width; the default value of {\small ``{\tt 0}''} indicates a single, inclusive bin.  The next two
inputs numerically specify the minimum and maximum cutoff values for inclusion in the histogram;  the lower boundary
defaults to {\small ``{\tt 0}''}, and the upper boundary defaults to {\small ``{\tt UNDEF}''}, which is interpreted to be indefinitely
large.  The fifth input is a integer list selector for the numerical label that should be assigned to each bin;  {\small ``{\tt 0}''}
(which is the default) labels each bin by its left-most (minimum, inclusive) value, while {\small ``{\tt 1}''} substitutes reporting
of the central bin value, and {\small ``{\tt 2}''} instead calls for the right-most (maximum, excluded) boundary value.
The last parameter input is a binary flag for governing the histogram output format; the default value of {\small ``{\tt 0}''}
produces standard tab-delimited text that is suitable for passing on to spreadsheet based statistics packages,
while the alternate specification of {\small ``{\tt 1}''} will yield a file employing the array input syntax favored by {\sc Mathematica}.

Additional helper scripts for interpreting the statistics of surviving groups of events and for automating the generation
of supervisory shell scripts to shepherd the processing of large batches of parallel analyses are available to interested
parties upon personal request.  No public documentation currently exists for these tools, although the curious are welcome
to explore and modify any provided code for adaptation to their own purposes, under the same terms~\cite{gnugpl} as the rest of the package.
In general, the author is quite open to private inquiries, suggestions, and requests for support by potential {\sc CutLHCO} users
on any topic regarding application of this software to the physics analysis of collider data selection cuts.

%%%%%%%%%%%%%%%%%%%%%%%%%%%%%%%%%%%%%%%%%%%%%%%%%%%

\section{Card File Case Studies\label{sct:cards}}

The most efficient way to communicate a working vocabulary of {\sc CutLHCO}
commands to the potential user is probably by direct example.  The general syntax tutorial 
from Section~(\ref{sct:syntax}), along with the summary of available parameters
in the document Appendix, are thus supplemented in the present section by five specific
case studies of card files designed for the replication of
actual LHC SUSY searches, three from the CMS collaboration, and two from ATLAS.
The coded instruction sets are are transcribed in {\small {\tt Typewriter}} font within
individually boxed panels, and a line-by-line deconstruction of the event selection
intent accompanies each such inset.  A growing library of several dozens of additional
LHC search card files is provided along with the online program distribution
package~\cite{cutlhco}.  Users who create interesting new search cards are encouraged to
submit them to the author for inclusion with future releases.

%%%%%%%%%%%%%%%%%%%%%%%%%%%%%%%%%%%%%%%%%%%%%%%%%%%

\subsection{CMS Hadronic Multijet Search}

\begin{card}
\centering
\ovalbox{%
\begin{minipage}{0.97\linewidth}
\vspace{3pt}
{\scriptsize
\begin{Verbatim}[numbers=left,numbersep=6pt]
****** cut_card.dat 2.0 ******
* ASA: CMS Hadronic Multijets
* CMS PAS SUS-11-003 & SUS-09-001
*** Object Reconstruction ****
OBJ_PHO = PTM:25, CUT:[0,0]
OBJ_ELE = PTM:10, CUT:[0,0]
OBJ_MUO = PTM:10, CUT:[0,0]
OBJ_JET = PTM:30
OBJ_JET_001 = SRC:+000, PTM:50, CUT:9
OBJ_JET_002 = SRC:+001, PTM:100, CUT:2
OBJ_JET_003 = SRC:+002, PRM:[0.0,2.5,1], CUT:1
OBJ_JET_004 = SRC:+001, PRM:3.0, CUT:[0,0]
OBJ_JET_005 = SRC:+001, FEM:0.9, CUT:[0,0]
****** Event Selection *******
EVT_MET_001 = JET:000
EVT_MET_002 = JET:001, CUT:100
EVT_MHT_001 = JET:001, CUT:375
EVT_RET_001 = NUM:002, DEN:001, CUT:[0.0,1.25]
******************************
\end{Verbatim}
} \vspace{-3pt} \end{minipage}}
\caption{CMS Hadronic Multijet Search~\cite{PAS-SUS-11-003,PAS-SUS-09-001}}
\end{card}

The first card file scenario to be documented here represents a search
by the CMS collaboration for SUSY signals in purely Hadronic multijet
events~\cite{PAS-SUS-11-003,PAS-SUS-09-001}.  The procedure specified 
in the two CMS references differs in various minor details, and the provided card file
thus represents something of a hybrid treatment.   Following a point of some persistent
interest to the author's research collaboration~\cite{Maxin:2011hy,Li:2011fu}, a high
jet multiplicity threshold requirement is enforced by the current card that does not
appear in the original CMS sources.  This example likewise omits a cut on the $\alpha_{\rm T}^{j,j}$
statistic of Eq.~(\ref{eq:alphat}), in keeping with the previously described difficulties
facing application of that measure to extreme multijet content.  The resulting selection
is appropriate, for example, for comparison against the data plotted in Figure~1b of
Ref.~\cite{PAS-SUS-11-003}, which shows the event distribution per jet count, up to a
maximum of 12, prior to application of the $\alpha_{\rm T}$ filter.

The object reconstruction begins in line {\footnotesize (5)} by discarding
events that possess photons meeting a 25~GeV minimal threshold on the transverse momentum $P_{\rm T}$.
Likewise, events with either electrons or muons carrying a $P_{\rm T}$ at or above 10~GeV are
subsequently eliminated. The remaining object specifications are dedicated
to the partitioning of hadronic jets.  Line {\footnotesize (8)} prescribes a soft zeroth order
cut of $P_{\rm T} \ge 30$~GeV on the transverse momentum of jets described in the input \dlhco event,
which will be inherited by all subsequent groupings.  A first positively indexed partition,
sourced from the prior zeroth order designation, is created from the jet object subset possessing
a harder transverse momentum selection of $P_{\rm T} \ge 50$~GeV, and events are rejected
that do not possess at least 9 such jets.  A second subdivision further requires
the presence of at least two leading hard jets with $P_{\rm T} \ge 100$~GeV.  The third jet
reconstruction index extends this sequential processing of the object chain to require
a pseudo-rapidity magnitude $\vert \eta \vert \le 2.5$ for the leading jet;
activation of the third position ``stop'' flag within the {\small {\tt PRM}} parameter forces 
the clustering of objects to terminate upon the first selection failure (in descending $P_{\rm T}$ order),
guaranteeing that a non-zero final count does indeed include the single hardest jet.
The final two groupings in lines {\footnotesize (12,13)} are sourced retroactively from the previously defined
$P_{\rm T} \ge 50$~GeV set in line {\footnotesize (9)}, and further isolate jets with a pseudo-rapidity
$\vert \eta \vert \ge 3.0$ or an electromagnetic fraction $\xi \ge 0.9$, rejecting the
event as a whole if any such surviving objects are identified.

The event selection opens in lines {\footnotesize (15,16)} with missing transverse energy definitions for two source
populations consisting, in turn, of all soft $P_{\rm T} \ge 30$~GeV jets and the harder $P_{\rm T} \ge 50$~GeV subset.
The latter of these groupings, which is guaranteed from before
to possess at least 9 objects with satisfactory $\eta$ and $\xi$ values, must yield a missing energy
of at least 100~GeV for the event to continue.  Subsequently, the same set of harder jets is
checked for its scalar transverse momentum sum $H_{\rm T}$, where a minimal value of 375~GeV is required.
Finally, an upper bound on the ratio of hard-to-soft jet missing energies
$( {\slashed{E}} {}_{\rm T}^{\rm hard} / {\slashed{E}} {}_{\rm T}^{\rm soft} ) \le 1.25$ is imposed.
If one wishes to reinstate the CMS $\alpha_{\rm T}^{j,j} \ge 0.55$ selection, this is easily accomplished by
addition of the line {\small ``{\tt EVT\_ALT\_001~=~JET:001,~CUT:0.55}''}.   Since the appropriate
{\small {\tt MET}} and {\small {\tt MHT}} factors may be reconstructed from the provided jets
they need not be supplied explicitly. 

%%%%%%%%%%%%%%%%%%%%%%%%%%%%%%%%%%%%%%%%%%%%%%%%%%%

\subsection{ATLAS Hadronic Multijet Search}
 
\begin{card}
\centering
\ovalbox{%
\begin{minipage}{0.97\linewidth}
\vspace{3pt}
{\scriptsize
\begin{Verbatim}[numbers=left,numbersep=6pt]
****** cut_card.dat 2.0 ******
* ATLAS Hadronic Multijets (7J80)
* ATLAS-CONF-2012-037
*** Object Reconstruction ****
OBJ_ALL = PRM:[0.0,4.5]
OBJ_ELE = PTM:20, PRM:[0.0,2.47]
OBJ_MUO = PTM:10, PRM:[0.0,2.4]
OBJ_JET = PTM:20, PRM:[0.0,2.8]
OBJ_LEP_001 = SRC:+000, EMT:+1
OBJ_JET_002 = SRC:+000, CMP:+001, CDR:0.2
OBJ_LEP_003 = SRC:+000, CMP:+002, EMT:-3, CDR:0.4, CUT:[0,0]
OBJ_JET_003 = SRC:+002, PTM:40
OBJ_JET_004 = SRC:+003, PTM:80, CUT:7
****** Event Selection *******
EVT_MHT_001 = JET:003
EVT_RHT_001 = NUM:000, DEN:001, CUT:4.0
******************************
\end{Verbatim}
} \vspace{-3pt} \end{minipage}}
\caption{ATLAS Hadronic Multijet Search~\cite{ATLAS-CONF-2012-037}}
\end{card}

The second card file example to be presented represents a parallel search by the ATLAS
collaboration for SUSY signals in purely Hadronic multijet events~\cite{ATLAS-CONF-2012-037}.
The specific selection strategy to be documented is referred to as ``7j80'', in reference
to the required jet count and transverse momentum per jet.
Though targeting an essentially identical event profile to that just described, the mechanism of
the adopted selections is distinct.

The object specification begins in line {\footnotesize (5)} with a global
restriction on the pseudo-rapidity magnitude $\vert \eta \vert \le 4.5$ that will be enforced
on all subsequently inherited objects.  The light lepton (electron,muon) populations
are then filtered according to transverse momentum $P_{\rm T} \ge (20,10)$~GeV and pseudo-rapidity
$\vert \eta \vert \le (2.47,2.40)$.  Likewise, the zeroth jet classification is restricted to
the object subset with $P_{\rm T} \ge 20$~GeV and $\vert \eta \vert \le 2.8$.  In line {\footnotesize (9)} 
a lepton reconstruction is sourced from the individual merged lepton flavors, and
filtered specifically to contain only electrons.  The inherited jet object set is then compared
against the prior electron grouping for angular proximity, and only those jets
that are isolated by $\Delta R \ge 0.2$ radians from the nearest electron are retained.
Next, a fresh assembly of leptons is gathered in line {\footnotesize (11)} that specifically exempts tau flavor
content.  It is evaluated for angular proximity against all jets surviving the filter defined previously in line
{\footnotesize (10)}, and the event is discarded as a whole if any well isolated leptons, having $\Delta R \ge 0.4$ with
respect to all jets, persist.  An intermediate transverse momentum cut of $P_{\rm T} \ge 40$~GeV is then
applied to the continued jets.  Immediately following is a further hard classification
of $P_{\rm T} \ge 80$~GeV, which must be survived by at least 7 jets for the event to proceed.

The event selection phase initiated in line {\footnotesize (15)} is rather compact, firstly defining a scalar transverse momentum sum
$H_{\rm T}^{\rm jets}$ for the intermediate jet classification, and secondly enforcing a relative
significance boundary of $ {\slashed{E}} {}_{\rm T}^{\rm all} / \surd[ H_{\rm T}^{\rm jets} ] \ge 4~{\rm GeV}^{1/2}$
on the missing transverse energy $ {\slashed{E}} {}_{\rm T}^{\rm all}$ computed from all objects
that were consistent with the primary pseudo-rapidity filter.

%%%%%%%%%%%%%%%%%%%%%%%%%%%%%%%%%%%%%%%%%%%%%%%%%%%

\subsection{CMS Dilepton and B-Tagged Jet Search}

\begin{card}
\centering
\ovalbox{%
\begin{minipage}{0.97\linewidth}
\vspace{3pt}
{\scriptsize
\begin{Verbatim}[numbers=left,numbersep=6pt]
****** cut_card.dat 2.0 ******
* CMS SSAF Dileptons and Bjets (SR1)
* CMS PAS SUS-11-020
*** Object Reconstruction ****
OBJ_ELE = PRM:[1.566,1.422]
OBJ_LEP = EMT:-3, PTM:10, PRM:[0.0,2.4]
OBJ_JET = PTM:40, PRM:[0.0,2.5], CUT:2
OBJ_LEP_001 = SRC:+000, SGN:+1, PTM:[UNDEF,20]
OBJ_LEP_002 = SRC:+000, SGN:-1, PTM:[UNDEF,20]
OBJ_LEP_003 = SRC:[+000,-001]
OBJ_LEP_004 = SRC:[+000,-002]
OBJ_LEP_005 = SRC:[+000,-001,-002]
OBJ_JET_001 = SRC:+000, HFT:1, CUT:2
OBJ_DIL_001 = LEP:003, DLS:-1, DLF:1, DMI:[76,106], CUT:[0,0]
OBJ_DIL_002 = LEP:004, DLS:-1, DLF:1, DMI:[76,106], CUT:[0,0]
OBJ_DIL_003 = LEP:005, DLS:+1, DLF:0, DMI:8, CUT:1
****** Event Selection *******
EVT_MET = CUT:30
EVT_MHT_001 = JET:000, CUT:80
******************************
\end{Verbatim}
} \vspace{-3pt} \end{minipage}}
\caption{CMS Dilepton and B-Tagged Jet Search~\cite{PAS-SUS-11-020}}
\end{card}

The next selection card deconstruction confronts an intricate CMS
search strategy formulated around the invariant mass of a same-sign, any-flavor
dilepton pair, and requiring the presence of jets with heavy flavor tagging~\cite{PAS-SUS-11-020}.
In particular, the example documents a region of the selection space referred to as ``SR1''.

The object reconstruction begins in line {\footnotesize (5)} by dismissing electrons that are located
within an angular dead region at the CMS barrel-endcap calorimeter seam corresponding to
the pseudo-rapidity magnitude \mbox{$ 1.422 < \vert \eta \vert < 1.566$} range.
Next, there is a specification of the zeroth lepton object set that
rejects intrusion by the tau, and places outer limits on the light flavor transverse
momentum $P_{\rm T} \ge 10$~GeV and pseudo-rapidity $\vert \eta \vert \le 2.4$.
Likewise, the zeroth jet classification, for which surviving events must provide at least
two matching objects, is bounded by $P_{\rm T} \ge 40$~GeV and $\vert \eta \vert \le 2.5$.
In lines {\footnotesize (8,9)} a pair of signed lepton groupings is sourced from the established zeroth
order set, one entirely positive and one entirely negative, each of which caps the transverse
momentum at $P_{\rm T} \le 20$~GeV.  Three additional lepton partitions
are then recursively defined by the selective veto of candidates from the zeroth sourcing pool
that also appear within one or either of the signed intermediate $P_{\rm T}$ pairs; the result is
a group of softer negative $P_{\rm T} \ge 10$~GeV leptons mixed with harder positive $P_{\rm T} > 20$~GeV
leptons, a similar group with precisely the reversed charge versus transverse momentum relationship,
and a group of mixed sign leptons that all respect the harder $P_{\rm T} > 20$~GeV classification.  A single
indexed jet partition is reduced from the zeroth classification in line {\footnotesize (13)} by specifying a loose selector 
on the heavy flavor tag, and events without at least 2 such tagged jets are are cut.  The final
three object reconstructions are of the dilepton variety.  The first two reject events containing
an opposite-sign same-flavor dilepton configuration that yields an invariant mass $M^{\ell,\ell}$, as in Eq.~(\ref{eq:minvab}),
within 15~GeV of $M_{\rm Z} \simeq 91$~GeV, if at least one member of the pair is of the harder
$P_{\rm T} > 20$~GeV object variety.  The final object specification line requires the presence
of at least 1 same-sign any-flavor dilepton pair with an invariant mass $M^{\ell,\ell} \ge 8$~GeV,
where each lepton is above the individual $P_{\rm T} > 20$~GeV threshold.

The event selection phase prescribed in lines {\footnotesize (18,19)} is quite straightforward, consisting of a cut
${\slashed{E}} {}_{\rm T}^{\rm all} \ge 30$~GeV on the missing energy composed from all primary event objects, and a cut
$ H_{\rm T}^{\rm jets} \ge 80$~GeV on the scalar transverse momentum sum for the zeroth assembly of jets.

%%%%%%%%%%%%%%%%%%%%%%%%%%%%%%%%%%%%%%%%%%%%%%%%%%%

\subsection{ATLAS Jets and Isolated Lepton Search}

\begin{card}
\centering
\ovalbox{%
\begin{minipage}{0.97\linewidth}
\vspace{3pt}
{\scriptsize
\begin{Verbatim}[numbers=left,numbersep=6pt]
****** cut_card.dat 2.0 ******
* ATLAS Jets and Lepton (3J1L)
* ATLAS-CONF-2012-041
*** Object Reconstruction ****
OBJ_ALL = PRM:[0.0,4.9]
OBJ_ELE = PTM:10, PRM:[0.0,2.47]
OBJ_MUO = PTM:10, PRM:[0.0,2.4]
OBJ_LEP_001 = SRC:+000, EMT:+1, PTM:25
OBJ_LEP_002 = SRC:+000, EMT:+2, PTM:20
OBJ_JET_002 = SRC:+000, CMP:+001, PTM:20, PRM:[0.0,4.5], CDR:0.2
OBJ_LEP_003 = SRC:[+001,+002], CMP:+002, CDR:0.4, CUT:[1,1]
OBJ_JET_003 = SRC:+002, PTM:25, PRM:[0.0,2.5], CUT:3
OBJ_LEP_004 = SRC:[+000,-003], EMT:-3, CUT:[0,0]
OBJ_JET_004 = SRC:+003, CUT:[3,UNDEF,-1]
OBJ_JET_005 = SRC:+003, PTM:80, CUT:[0,3]
OBJ_JET_006 = SRC:+005, PTM:100, CUT:1
****** Event Selection *******
EVT_MET = CUT:250
EVT_MHT_001 = LEP:003, JET:004
EVT_MEF_001 = MET:000, MHT:001
EVT_REF_001 = NUM:000, DEN:001, CUT:0.3
EVT_LTM_001 = LEP:003, MET:000, CUT:100
EVT_MHT_002 = LEP:003, JET:003
EVT_MEF_002 = MET:000, MHT:002, CUT:1200
******************************
\end{Verbatim}
} \vspace{-3pt} \end{minipage}}
\caption{ATLAS Jets and Isolated Lepton Search~\cite{ATLAS-CONF-2012-041}}
\end{card}

The second ATLAS collaboration card file scenario to be documented represents a
multi-level selection employed to search for SUSY in final states with jets, missing transverse
momentum, and a single isolated lepton~\cite{ATLAS-CONF-2012-041}.
An important role is played in this procedure by the transverse mass statistic
of Eq.~(\ref{eq:mt}).  The example provided corresponds specifically to the 3-jet selection sub-strategy.
Most ATLAS SUSY searches share a common core of generic low-level
object reconstruction filters, and the present card file specification
does indeed bear several initial structural similarities to that featured
in the previous ATLAS case study.

The object reconstruction begins in line {\footnotesize (5)} by enforcing a primary upper
limit on the pseudo-rapidity magnitude $\vert \eta \vert \le 4.9$ of all input objects.
The light lepton (electron,muon) populations are then filtered according to
transverse momentum $P_{\rm T} \ge 10$~GeV and pseudo-rapidity
$\vert \eta \vert \le (2.47,2.40)$.  In the next pair of lines a slightly harder
variant of each of these two distinct lepton flavor populations is indexed relative
to the implicit zeroth lepton composite, raising the lower limits
on transverse momentum to 25~GeV and 20~GeV, respectively.
An initial jet classification in line {\footnotesize (10)} is likewise sourced
from its corresponding inclusive zeroth grouping, enforcing $P_{\rm T} \ge 20$~GeV and
$\vert \eta \vert \le 4.5$ limits, and rejecting objects that are poorly isolated ($\Delta R < 0.2$ radians)
from members of the preceding electronic lepton partition defined in line {\footnotesize (8)}.  The electron and
muon forks are subsequently rejoined into a unified object source for the next instruction line, which filters against a minimal
isolation requirement of $\Delta R \ge 0.4$ radians from elements of the just established jet reconstruction,
and rejects events that do not afford 1, and only 1, compliant object.  After this, a slightly
harder and substantially more central jet partition is sourced from its direct numerical predecessor,
with $P_{\rm T} \ge 25$~GeV and $\vert \eta \vert \le 2.5$ limits, requiring a minimum of three
such jets for event continuation.  Line {\footnotesize (13)} marks a final return to the leptonic analysis,
prescribing a cut on events where any non-tau flavored object from the zeroth
classification proved too soft, forward or poorly isolated to pass through the remainder of the
described selection cascade.  The last three object reconstruction commands are additional
jet assembly protocols.  Picking up where the prior jet selection left off, the first of
these lines imposes a redundant lower bound of 3 on the surviving jet content, but sets
a value of {\small ``{\tt -1}''} for the third position ``clip'' input to the
{\small {\tt CUT}} parameter, which prevents any jets beyond the minimally required count
from appearing within the indexed object partition.  The next instruction, which rejects events
possessing more than 3 hard jets with $P_{\rm T} \ge 80$~GeV, returns for its sourcing to the classification
defined in line {\footnotesize (12)} rather than perpetuating the sequential filtering pattern; this discontinuity
is essential, given that the immediately prior classification was capped at a 3 jet maximum.
Resuming a stepwise object flow, the final jet specification requires the event to
contain at least 1 jet meeting the $P_{\rm T} \ge 100$~GeV threshold. 

The event selection phase opens in line {\footnotesize (18)} by placing a lower bound on the inclusive
missing energy ${\slashed{E}} {}_{\rm T}^{\rm all} \ge 250$~GeV.  Next, an indexed scalar
transverse energy sum specification ${H} {}_{\rm T}^{\ell,{\rm 3jet}} $
is created from back-references to the single surviving lepton and the capped partition
of 3 jets, as defined in lines {\footnotesize (11,14)}, respectively.  Likewise, an indexed effective mass $M_{\rm T}^{\rm eff}$
specification is established as a sum of the prior two quantities.  Subsequently, a limit
${\slashed{E}} {}_{\rm T}^{\rm all} / ( {\slashed{E}} {}_{\rm T}^{\rm all} + {H} {}_{\rm T}^{\ell,{\rm 3jet}}) \ge 0.3$
is placed on the corresponding missing energy to effective mass ratio.
Proceeding on, a cut $M_{\rm T}^{\ell,{\slashed{E}}} \ge 100$~GeV is instituted in line {\footnotesize (22)}
on the transverse mass composed from the isolated lepton and the event missing momentum vector.  In the next-to-final
instruction, a more inclusive transverse energy sum ${H} {}_{\rm T}^{\ell,{\rm jets}} $ is defined that
incorporates the lepton and all jets (not only the leading 3) within the
$P_{\rm T} \ge 25$~GeV and $\vert \eta \vert \le 2.5$ reconstruction from line {\footnotesize (12)}.  Finally, a lower bound
is imposed on the effective mass $( {\slashed{E}} {}_{\rm T}^{\rm all} + {H} {}_{\rm T}^{\ell,{\rm jets}}) \ge 1,200$~GeV
formed by combining the previous scalar sum with the event missing energy.

%%%%%%%%%%%%%%%%%%%%%%%%%%%%%%%%%%%%%%%%%%%%%%%%%%%

\subsection{CMS Razor Variable Supersymmetry Search}

\begin{card}
\centering
\ovalbox{%
\begin{minipage}{0.97\linewidth}
\vspace{3pt}
{\scriptsize
\begin{Verbatim}[numbers=left,numbersep=6pt]
****** cut_card.dat 2.0 ******
* CMS Razor ELE Box (SR6)
* CMS PAS SUS-12-005
*** Object Reconstruction ****
OBJ_ELE = PRM:[1.566,1.422]
OBJ_MUO = PRM:[0.0,2.4]
OBJ_LEP = EMT:-3, PTM:10, PRM:[0.0,2.5]
OBJ_JET = PTM:60, PRM:[0.0,3.0]
OBJ_LEP_001 = SRC:+000, EMT:+1
OBJ_LEP_002 = SRC:+000, EMT:+2
OBJ_LEP_003 = SRC:+002, ETR:[0.00,0.27], PRM:[0.0,2.1]
# OBJ_LEP_004 = SRC:+003, PTM:12, CUT:[0,0]
# OBJ_LEP_005 = SRC:+001, PTM:20, CUT:[0,0], ANY:004
OBJ_LEP_006 = SRC:+003, CUT:[0,0]
OBJ_LEP_007 = SRC:+002, PTM:15, CUT:[0,0]
OBJ_LEP_008 = SRC:+002, CUT:[0,1], ANY:[006,007]
# OBJ_LEP_009 = SRC:+001, PTM:20, CUT:[0,0]
OBJ_LEP_010 = SRC:+001, CUT:[0,1], ANY:009
OBJ_LEP_011 = SRC:+003, PTM:12, CUT:[0,0]
OBJ_LEP_012 = SRC:+001, PTM:20, CUT:1
****** Event Selection *******
EVT_JRM_001 = LEP:000, JET:000, CUT:[450,1000]
EVT_ALR_001 = LEP:000, JET:000, MET:000, CUT:[0.30,0.50]
******************************
\end{Verbatim}
} \vspace{-3pt} \end{minipage}}
\caption{CMS Razor Variable Supersymmetry Search~\cite{Chatrchyan:2011ek,PAS-SUS-12-005}}
\end{card}

The last card file scenario to be documented corresponds 
to a CMS search for SUSY using the razor variables~\cite{Chatrchyan:2011ek,PAS-SUS-12-005}.
Additional information on the active lepton reconstruction strategy and isolation requirements may be found in Ref.~\cite{PAS-EWK-10-002}.
The selection strategy to be highlighted hierarchically divides the harvested event content
into one of several ``boxes'' according to the lepton (or dilepton) flavor content, or the
lack thereof.  To keep the various groupings disjoint, events satisfying the
requirements of multiple boxes are uniquely sequestered upon their
first match against an ordered sequence of clustered selection criteria.  Conversely,
the assignment of an event into one of the latter classification stages requires demonstrating
that none of the prior available box qualifications were successfully matched.  In general, 
each box will possess multiple positive attributes that must all be satisfied for inclusion
of the event; therefore, a negation of this inclusion is facilitated by falsifying any
single member of the composite list of properties.  The required logic may be implemented
within the card file description by application of the {\small {\tt ANY}} parameter to
link related filters. 

The example provided corresponds specifically to signal region 6 of the single isolated
electron box from Ref.~\cite{PAS-SUS-12-005}. The object reconstruction begins in line
{\footnotesize (5)} with the exclusion of electrons located within the pseudo-rapidity range
corresponding to the barrel-endcap detector gap.  Next, muons are required to be contained 
within $\vert \eta \vert \le 2.4$, and a combined electron-muon lepton classification is
composed with transverse momentum $P_{\rm T} \ge 10$~GeV and the slightly weaker
pseudo-rapidity limit of $\vert \eta \vert \le 2.5$.  Similarly, the zeroth jet classification
is defined with $P_{\rm T} \ge 60$~GeV and $\vert \eta \vert \le 3.0$.  In lines \mbox{\footnotesize (9--11)}
the electron and muon objects are reseparated, and a tightened muon object grouping is instantiated
with an explicit bound on the transverse momentum isolation ratio $\zeta \le 0.27$ 
and a more central track orientation $\vert \eta \vert \le 2.1$.  Lines {\footnotesize (12,13)}
are the first content negation grouping to be tethered by the {\small {\tt ANY}} keyword.  The
two members of this construct are actually made redundant by subsequent instructions, and,
although their content is retained for pedagogical purposes, the leading comment markers {\small {\tt \#}}
will suppress interpretation of the associated commands.  In concert, events would be retained by this
pair of instructions that lacked either a tight muon with $P_{\rm T} \ge 12$~GeV or a hard electron with
$P_{\rm T} \ge 20$~GeV, both of which are mandated by the first box definition.  The next three lines similarly retain
events disqualified from inclusion in the second box, which is defined by the presence of at least
one tight muon, at least one harder muon with $P_{\rm T} \ge 15$~GeV, and at least two muons overall.
Lines {\footnotesize (17,18)}, one of which is again duplicative, would act collectively to exclude
events that lack either a single electron with $P_{\rm T} \ge 20$~GeV or two electrons overall.
Line {\footnotesize (19)} constitutes a negation of the fourth lepton box criterion, namely the
presence of at least one tight muon with $P_{\rm T} \ge 12$~GeV.  Since this is a precise duplication
of line {\footnotesize (12)}, which is logically coupled to line {\footnotesize (13)}, there is no
need to explicitly execute either of the prior tests.  The final object reconstruction command grants
otherwise successful events admission into the fifth leptonic box according to the presence of at least
one hard electron with $P_{\rm T} \ge 20$~GeV.  Surviving objects will thus necessarily fail the test embodied
in line {\footnotesize (17)}, which may consequently be omitted from its cluster.  The same holds true for
line {\footnotesize (13)}, although its action was already blocked for independent reasons. 

The event selection phase prescribed in lines {\footnotesize (22,23)} of this card is quite simple, isolating a rectangular area of 
the razor plane with \mbox{$450~{\rm GeV} \le M_{\rm R}^{j,j} \le 1000~{\rm GeV}$} and \mbox{$0.30 \le \alpha_{\rm R}^{j,j} \le 0.50$}
that is sourced from the inclusive missing transverse energy and the full retained classifications of leptons and jets.

%%%%%%%%%%%%%%%%%%%%%%%%%%%%%%%%%%%%%%%%%%%%%%%%%%%%%%%%%%%%%%%%%%%%%%%%%%%%

\section{Afterword}

A new computer program named {\sc CutLHCO} has been introduced, whose function is the implementation
of generic data selection cuts on collider event specification files in the standardized \dlhco format.
This software is intended to fill an open market niche for a lightweight yet flexible ``consumer-level''
alternative to the {\sc Root}~\cite{Brun:1997pa} data analysis framework.  The primary envisioned application is as 
a filter on output produced by the {\sc PGS4}~\cite{PGS4} and {\sc Delphes}~\cite{Ovyn:2009tx} detector simulations, which
are themselves lightweight alternatives to the {\sc Geant4}~\cite{Agostinelli:2002hh} based solutions favored by the large
LHC experiments.  All process control instructions are provided via a compact and powerful
card file input syntax that efficiently facilitates the reasonable approximation of most
event selection strategies and specialized discovery statistics commonly employed by the
CMS and ATLAS collaborations.  The structure, function, invocation and usage of the most recent
{\sc CutLHCO 2.0} program version has been documented thoroughly, including a detailed deconstruction
of several example card file specifications.  The associated software is now
available for free public download~\cite{cutlhco}.

%%%%%%%%%%%%%%%%%%%%%%%%%%%%%%%%%%%%%%%%%%%%%%%%%%%%%%%%%%%%%%%%%%%%%%%%%%%%

\begin{acknowledgments}
I express gratitude to Dimitri V. Nanopoulos, Tianjun Li, and James A. Maxin,
colleagues in an ongoing phenomenological particle physics study that occasioned
development of the software described in this article. 
I thank Sam Houston State University for providing the high performance computing
resources used to develop, refine and employ this software.
This research was also supported in part by the Sam Houston State University
2011-2012 Enhancement Research Grant program.
\end{acknowledgments}

%%%%%%%%%%%%%%%%%%%%%%%%%%%%%%%%%%%%%%%%%%%%%%%%%%%%%%%%%%%%%%%%%%%%%%%%%%%%

\makeatletter
\close@column@grid
\onecolumngrid
\vspace{8pt}
\begin{center}
\includegraphics[width=0.8\textwidth]{vectorian_dot_net_flourish.eps}
\end{center}
\close@column@grid
\clearpage
\twocolumngrid
\makeatother

%%%%%%%%%%%%%%%%%%%%%%%%%%%%%%%%%%%%%%%%%%%%%%%%%%%%%%%%%%%%%%%%%%%%%%%%%%%%

\def\bibsection{\section*{References}} 

%%%%%%%%%%%%%%%%%%%%%%%%%%%%%%%%%%%%%%%%%%%%%%%%%%%%%%%%%%%%%%%%%%%%%%%%%%%%

\bibliography{bibliography}

\begin{thebibliography}{39}
\expandafter\ifx\csname natexlab\endcsname\relax\def\natexlab#1{#1}\fi
\expandafter\ifx\csname bibnamefont\endcsname\relax
  \def\bibnamefont#1{#1}\fi
\expandafter\ifx\csname bibfnamefont\endcsname\relax
  \def\bibfnamefont#1{#1}\fi
\expandafter\ifx\csname citenamefont\endcsname\relax
  \def\citenamefont#1{#1}\fi
\expandafter\ifx\csname url\endcsname\relax
  \def\url#1{\texttt{#1}}\fi
\expandafter\ifx\csname urlprefix\endcsname\relax\def\urlprefix{URL }\fi
\providecommand{\bibinfo}[2]{#2}
\providecommand{\eprint}[2][]{\url{#2}}

\bibitem[{\citenamefont{Mangano et~al.}(2003)\citenamefont{Mangano, Moretti,
  Piccinini, Pittau, and Polosa}}]{Mangano:2002ea}
\bibinfo{author}{\bibfnamefont{M.~L.} \bibnamefont{Mangano}},
  \bibinfo{author}{\bibfnamefont{M.}~\bibnamefont{Moretti}},
  \bibinfo{author}{\bibfnamefont{F.}~\bibnamefont{Piccinini}},
  \bibinfo{author}{\bibfnamefont{R.}~\bibnamefont{Pittau}}, \bibnamefont{and}
  \bibinfo{author}{\bibfnamefont{A.~D.} \bibnamefont{Polosa}},
  {``}\bibinfo{title}{{ALPGEN, a generator for hard multiparton processes in
  hadronic collisions}},{''} \bibinfo{journal}{JHEP}
  \textbf{\bibinfo{volume}{0307}}, \bibinfo{pages}{001} (\bibinfo{year}{2003}),
  \eprint{hep-ph/0206293}

\bibitem[{\citenamefont{Stelzer and Long}(1994)}]{Stelzer:1994ta}
\bibinfo{author}{\bibfnamefont{T.}~\bibnamefont{Stelzer}} \bibnamefont{and}
  \bibinfo{author}{\bibfnamefont{W.~F.} \bibnamefont{Long}},
  {``}\bibinfo{title}{{Automatic generation of tree level helicity
  amplitudes}},{''} \bibinfo{journal}{Comput. Phys. Commun.}
  \textbf{\bibinfo{volume}{81}}, \bibinfo{pages}{357} (\bibinfo{year}{1994}),
  \eprint{hep-ph/9401258}

\bibitem[{\citenamefont{Alwall et~al.}(2007)}]{Alwall:2007st}
\bibinfo{author}{\bibfnamefont{J.}~\bibnamefont{Alwall}} \bibnamefont{et~al.},
  {``}\bibinfo{title}{{MadGraph/MadEvent v4: The New Web Generation}},{''}
  \bibinfo{journal}{JHEP} \textbf{\bibinfo{volume}{09}}, \bibinfo{pages}{028}
  (\bibinfo{year}{2007}), \eprint{0706.2334}

\bibitem[{\citenamefont{Sjostrand et~al.}(2006)\citenamefont{Sjostrand, Mrenna,
  and Skands}}]{Sjostrand:2006za}
\bibinfo{author}{\bibfnamefont{T.}~\bibnamefont{Sjostrand}},
  \bibinfo{author}{\bibfnamefont{S.}~\bibnamefont{Mrenna}}, \bibnamefont{and}
  \bibinfo{author}{\bibfnamefont{P.~Z.} \bibnamefont{Skands}},
  {``}\bibinfo{title}{{PYTHIA 6.4 Physics and Manual}},{''}
  \bibinfo{journal}{JHEP} \textbf{\bibinfo{volume}{05}}, \bibinfo{pages}{026}
  (\bibinfo{year}{2006}), \eprint{hep-ph/0603175}

\bibitem[{\citenamefont{Corcella et~al.}(2001)\citenamefont{Corcella, Knowles,
  Marchesini, Moretti et~al.}}]{Corcella:2000bw}
\bibinfo{author}{\bibfnamefont{G.}~\bibnamefont{Corcella}},
  \bibinfo{author}{\bibfnamefont{I.}~\bibnamefont{Knowles}},
  \bibinfo{author}{\bibfnamefont{G.}~\bibnamefont{Marchesini}},
  \bibinfo{author}{\bibfnamefont{S.}~\bibnamefont{Moretti}},
  \bibnamefont{et~al.}, {``}\bibinfo{title}{{HERWIG 6: An Event generator for
  hadron emission reactions with interfering gluons (including supersymmetric
  processes)}},{''} \bibinfo{journal}{JHEP} \textbf{\bibinfo{volume}{0101}},
  \bibinfo{pages}{010} (\bibinfo{year}{2001}), \eprint{hep-ph/0011363}

\bibitem[{\citenamefont{Gleisberg et~al.}(2009)\citenamefont{Gleisberg, Hoeche,
  Krauss, Schonherr, Schumann et~al.}}]{Gleisberg:2008ta}
\bibinfo{author}{\bibfnamefont{T.}~\bibnamefont{Gleisberg}},
  \bibinfo{author}{\bibfnamefont{S.}~\bibnamefont{Hoeche}},
  \bibinfo{author}{\bibfnamefont{F.}~\bibnamefont{Krauss}},
  \bibinfo{author}{\bibfnamefont{M.}~\bibnamefont{Schonherr}},
  \bibinfo{author}{\bibfnamefont{S.}~\bibnamefont{Schumann}},
  \bibnamefont{et~al.}, {``}\bibinfo{title}{{Event generation with SHERPA
  1.1}},{''} \bibinfo{journal}{JHEP} \textbf{\bibinfo{volume}{0902}},
  \bibinfo{pages}{007} (\bibinfo{year}{2009}), \eprint{0811.4622}

\bibitem[{\citenamefont{Agostinelli et~al.}(2003)}]{Agostinelli:2002hh}
\bibinfo{author}{\bibfnamefont{S.}~\bibnamefont{Agostinelli}}
  \bibnamefont{et~al.} (\bibinfo{collaboration}{GEANT4}),
  {``}\bibinfo{title}{{GEANT4: A Simulation toolkit}},{''}
  \bibinfo{journal}{Nucl.Instrum.Meth.} \textbf{\bibinfo{volume}{A506}},
  \bibinfo{pages}{250} (\bibinfo{year}{2003})

\bibitem[{\citenamefont{Conway et~al.}(2009)}]{PGS4}
\bibinfo{author}{\bibfnamefont{J.}~\bibnamefont{Conway}} \bibnamefont{et~al.},
  {``}\bibinfo{title}{PGS4: Pretty Good (Detector) Simulation},{''}
  (\bibinfo{year}{2009}),
  \urlprefix\url{http://physics.ucdavis.edu/~conway/research/software/pgs/pgs4-general.htm}

\bibitem[{\citenamefont{Ovyn et~al.}(2009)\citenamefont{Ovyn, Rouby, and
  Lemaitre}}]{Ovyn:2009tx}
\bibinfo{author}{\bibfnamefont{S.}~\bibnamefont{Ovyn}},
  \bibinfo{author}{\bibfnamefont{X.}~\bibnamefont{Rouby}}, \bibnamefont{and}
  \bibinfo{author}{\bibfnamefont{V.}~\bibnamefont{Lemaitre}},
  {``}\bibinfo{title}{{DELPHES, a framework for fast simulation of a generic
  collider experiment}},{''} (\bibinfo{year}{2009}), \eprint{0903.2225}

\bibitem[{\citenamefont{Brun and Rademakers}(1997)}]{Brun:1997pa}
\bibinfo{author}{\bibfnamefont{R.}~\bibnamefont{Brun}} \bibnamefont{and}
  \bibinfo{author}{\bibfnamefont{F.}~\bibnamefont{Rademakers}},
  {``}\bibinfo{title}{{ROOT: An object oriented data analysis framework}},{''}
  \bibinfo{journal}{Nucl.Instrum.Meth.} \textbf{\bibinfo{volume}{A389}},
  \bibinfo{pages}{81} (\bibinfo{year}{1997})

\bibitem[{\citenamefont{Conte et~al.}(2012)\citenamefont{Conte, Fuks, and
  Serret}}]{Conte:2012fm}
\bibinfo{author}{\bibfnamefont{E.}~\bibnamefont{Conte}},
  \bibinfo{author}{\bibfnamefont{B.}~\bibnamefont{Fuks}}, \bibnamefont{and}
  \bibinfo{author}{\bibfnamefont{G.}~\bibnamefont{Serret}},
  {``}\bibinfo{title}{{MadAnalysis 5, a user-friendly framework for collider
  phenomenology}},{''} (\bibinfo{year}{2012}), \eprint{1206.1599}

\bibitem[{lhc(2007)}]{lhcowiki}
{``}\bibinfo{title}{LHC Olympics Wiki},{''} (\bibinfo{year}{2007}),
  \urlprefix\url{http://www.jthaler.net/olympicswiki/}

\bibitem[{\citenamefont{Li et~al.}(2011{\natexlab{a}})\citenamefont{Li, Maxin,
  Nanopoulos, and Walker}}]{Maxin:2011hy}
\bibinfo{author}{\bibfnamefont{T.}~\bibnamefont{Li}},
  \bibinfo{author}{\bibfnamefont{J.~A.} \bibnamefont{Maxin}},
  \bibinfo{author}{\bibfnamefont{D.~V.} \bibnamefont{Nanopoulos}},
  \bibnamefont{and} \bibinfo{author}{\bibfnamefont{J.~W.}
  \bibnamefont{Walker}}, {``}\bibinfo{title}{{The Ultrahigh jet multiplicity
  signal of stringy no-scale $\cal{F}$-$SU(5)$ at the $\sqrt{s}= 7$ TeV
  LHC}},{''} \bibinfo{journal}{Phys.Rev.} \textbf{\bibinfo{volume}{D84}},
  \bibinfo{pages}{076003} (\bibinfo{year}{2011}{\natexlab{a}}),
  \eprint{1103.4160}

\bibitem[{\citenamefont{Li et~al.}(2011{\natexlab{b}})\citenamefont{Li, Maxin,
  Nanopoulos, and Walker}}]{Li:2011fu}
\bibinfo{author}{\bibfnamefont{T.}~\bibnamefont{Li}},
  \bibinfo{author}{\bibfnamefont{J.~A.} \bibnamefont{Maxin}},
  \bibinfo{author}{\bibfnamefont{D.~V.} \bibnamefont{Nanopoulos}},
  \bibnamefont{and} \bibinfo{author}{\bibfnamefont{J.~W.}
  \bibnamefont{Walker}}, {``}\bibinfo{title}{{Has SUSY Gone Undetected in 9-jet
  Events? A Ten-Fold Enhancement in the LHC Signal Efficiency}},{''}
  (\bibinfo{year}{2011}{\natexlab{b}}), \eprint{1108.5169}

\bibitem[{\citenamefont{Li et~al.}(2011{\natexlab{c}})\citenamefont{Li, Maxin,
  Nanopoulos, and Walker}}]{Li:2011av}
\bibinfo{author}{\bibfnamefont{T.}~\bibnamefont{Li}},
  \bibinfo{author}{\bibfnamefont{J.~A.} \bibnamefont{Maxin}},
  \bibinfo{author}{\bibfnamefont{D.~V.} \bibnamefont{Nanopoulos}},
  \bibnamefont{and} \bibinfo{author}{\bibfnamefont{J.~W.}
  \bibnamefont{Walker}}, {``}\bibinfo{title}{{Profumo di SUSY: Suggestive
  Correlations in the ATLAS and CMS High Jet Multiplicity Data}},{''}
  (\bibinfo{year}{2011}{\natexlab{c}}), \eprint{1111.4204}

\bibitem[{\citenamefont{Li et~al.}(2012)\citenamefont{Li, Maxin, Nanopoulos,
  and Walker}}]{Li:2012tr}
\bibinfo{author}{\bibfnamefont{T.}~\bibnamefont{Li}},
  \bibinfo{author}{\bibfnamefont{J.~A.} \bibnamefont{Maxin}},
  \bibinfo{author}{\bibfnamefont{D.~V.} \bibnamefont{Nanopoulos}},
  \bibnamefont{and} \bibinfo{author}{\bibfnamefont{J.~W.}
  \bibnamefont{Walker}}, {``}\bibinfo{title}{{Chanel ${\rm N^o5}$ (${\rm
  fb}^{-1}$): The Sweet Fragrance of SUSY}},{''} (\bibinfo{year}{2012}),
  \eprint{1205.3052}

\bibitem[{gnu(2007)}]{gnugpl}
{``}\bibinfo{title}{GNU General Public License, Version 3},{''}
  (\bibinfo{year}{2007}), \urlprefix\url{http://www.gnu.org/licenses/gpl.html}

\bibitem[{\citenamefont{Walker}(2012)}]{cutlhco}
\bibinfo{author}{\bibfnamefont{J.~W.} \bibnamefont{Walker}},
  {``}\bibinfo{title}{CutLHCO 2.0},{''} (\bibinfo{year}{2012}),
  \urlprefix\url{http://www.joelwalker.net/code/cut_lhco.tar.gz}

\bibitem[{PAS(2009)}]{PAS-SUS-09-001}
{``}\bibinfo{title}{{Search strategy for exclusive multi-jet events from
  supersymmetry at CMS}},{''} (\bibinfo{year}{2009}), \bibinfo{note}{{CMS PAS
  SUS-09-001}}

\bibitem[{PAS(2010)}]{PAS-EWK-10-002}
{``}\bibinfo{title}{{Measurements of Inclusive W and Z Cross Sections in pp
  Collisions at $\sqrt{s} = 7$~TeV}},{''} (\bibinfo{year}{2010}),
  \bibinfo{note}{{CMS PAS EWK-10-002}}

\bibitem[{\citenamefont{ATLAS}(2012)}]{ATLAS-CONF-2012-041}
\bibinfo{author}{\bibnamefont{ATLAS}}, {``}\bibinfo{title}{{Further search for
  supersymmetry at $\sqrt{s}=7$ TeV in final states with jets, missing
  transverse momentum and one isolated lepton}},{''} (\bibinfo{year}{2012}),
  \bibinfo{note}{{ATLAS-CONF-2012-041}}

\bibitem[{ATL(2011)}]{ATLAS-CONF-2011-130}
{``}\bibinfo{title}{{Search for supersymmetry in pp collisions at $\sqrt{s}$ =
  7 TeV in final states with missing transverse momentum, b-jets and one lepton
  with the ATLAS detector}},{''} (\bibinfo{year}{2011}),
  \bibinfo{note}{{ATLAS-CONF-2011-130}}

\bibitem[{\citenamefont{Aad et~al.}(2012)}]{ATLAS:2011ad}
\bibinfo{author}{\bibfnamefont{G.}~\bibnamefont{Aad}} \bibnamefont{et~al.}
  (\bibinfo{collaboration}{ATLAS Collaboration}), {``}\bibinfo{title}{{Search
  for supersymmetry in final states with jets, missing transverse momentum and
  one isolated lepton in $\sqrt{s} = 7$~TeV pp collisions using 1 ${\rm
  fb}^{-1}$ of ATLAS data}},{''} \bibinfo{journal}{Phys.Rev.}
  \textbf{\bibinfo{volume}{D85}}, \bibinfo{pages}{012006}
  (\bibinfo{year}{2012}), \eprint{1109.6606}

\bibitem[{PAS(2011{\natexlab{a}})}]{PAS-SUS-11-015}
{``}\bibinfo{title}{{Search for supersymmetry in pp collisions at $\sqrt{s} =
  7$~TeV in events with a single lepton, jets, and missing transverse
  momentum}},{''} (\bibinfo{year}{2011}{\natexlab{a}}), \bibinfo{note}{{CMS PAS
  SUS-11-015}}

\bibitem[{\citenamefont{Jackson}(2000)}]{kinematics}
\bibinfo{author}{\bibfnamefont{J.}~\bibnamefont{Jackson}},
  {``}\bibinfo{title}{{Kinematics}},{''} (\bibinfo{year}{2000}),
  \bibinfo{note}{particle Data Group Review},
  \urlprefix\url{http://pdg.lbl.gov/2008/reviews/kinemarpp.pdf}

\bibitem[{PAS(2012{\natexlab{a}})}]{PAS-SUS-12-002}
{``}\bibinfo{title}{{Search for Supersymmetry in hadronic Final states using
  $M_{T2}$ with the CMS detector at $\sqrt{s} = 7$~TeV}},{''}
  (\bibinfo{year}{2012}{\natexlab{a}}), \bibinfo{note}{{CMS PAS SUS-12-002}}

\bibitem[{\citenamefont{Lester and Summers}(1999)}]{Lester:1999tx}
\bibinfo{author}{\bibfnamefont{C.}~\bibnamefont{Lester}} \bibnamefont{and}
  \bibinfo{author}{\bibfnamefont{D.}~\bibnamefont{Summers}},
  {``}\bibinfo{title}{{Measuring masses of semiinvisibly decaying particles
  pair produced at hadron colliders}},{''} \bibinfo{journal}{Phys.Lett.}
  \textbf{\bibinfo{volume}{B463}}, \bibinfo{pages}{99} (\bibinfo{year}{1999}),
  \eprint{hep-ph/9906349}

\bibitem[{\citenamefont{Cho et~al.}(2008)\citenamefont{Cho, Choi, Kim, and
  Park}}]{Cho:2007dh}
\bibinfo{author}{\bibfnamefont{W.~S.} \bibnamefont{Cho}},
  \bibinfo{author}{\bibfnamefont{K.}~\bibnamefont{Choi}},
  \bibinfo{author}{\bibfnamefont{Y.~G.} \bibnamefont{Kim}}, \bibnamefont{and}
  \bibinfo{author}{\bibfnamefont{C.~B.} \bibnamefont{Park}},
  {``}\bibinfo{title}{{Measuring superparticle masses at hadron collider using
  the transverse mass kink}},{''} \bibinfo{journal}{JHEP}
  \textbf{\bibinfo{volume}{0802}}, \bibinfo{pages}{035} (\bibinfo{year}{2008}),
  \eprint{0711.4526}

\bibitem[{\citenamefont{Bayatian et~al.}(2007)}]{Ball:2007zza}
\bibinfo{author}{\bibfnamefont{G.}~\bibnamefont{Bayatian}} \bibnamefont{et~al.}
  (\bibinfo{collaboration}{CMS Collaboration}), {``}\bibinfo{title}{{CMS
  technical design report, volume II: Physics performance}},{''}
  \bibinfo{journal}{J.Phys.G} \textbf{\bibinfo{volume}{G34}},
  \bibinfo{pages}{995} (\bibinfo{year}{2007})

\bibitem[{\citenamefont{Chatrchyan
  et~al.}(2012{\natexlab{a}})}]{Chatrchyan:2012qka}
\bibinfo{author}{\bibfnamefont{S.}~\bibnamefont{Chatrchyan}}
  \bibnamefont{et~al.} (\bibinfo{collaboration}{CMS Collaboration}),
  {``}\bibinfo{title}{{Search for physics beyond the standard model in events
  with a Z boson, jets, and missing transverse energy in pp collisions at
  $\sqrt{s} = 7$~TeV}},{''} (\bibinfo{year}{2012}{\natexlab{a}}),
  \eprint{1204.3774}

\bibitem[{\citenamefont{Chatrchyan
  et~al.}(2012{\natexlab{b}})}]{Chatrchyan:2011ek}
\bibinfo{author}{\bibfnamefont{S.}~\bibnamefont{Chatrchyan}}
  \bibnamefont{et~al.} (\bibinfo{collaboration}{CMS Collaboration}),
  {``}\bibinfo{title}{{Inclusive search for squarks and gluinos in pp
  collisions at $\sqrt{s} = 7$~TeV}},{''} \bibinfo{journal}{Phys.Rev.}
  \textbf{\bibinfo{volume}{D85}}, \bibinfo{pages}{012004}
  (\bibinfo{year}{2012}{\natexlab{b}}), \eprint{1107.1279}

\bibitem[{PAS(2012{\natexlab{b}})}]{PAS-SUS-12-005}
{``}\bibinfo{title}{{Search for Supersymmetry with the Razor Variables}},{''}
  (\bibinfo{year}{2012}{\natexlab{b}}), \bibinfo{note}{{CMS PAS SUS-12-005}}

\bibitem[{\citenamefont{Rogan}(2010)}]{Rogan:2010kb}
\bibinfo{author}{\bibfnamefont{C.}~\bibnamefont{Rogan}},
  {``}\bibinfo{title}{{Kinematical variables towards new dynamics at the
  LHC}},{''} (\bibinfo{year}{2010}), \eprint{1006.2727}

\bibitem[{PAS(2008)}]{PAS-SUS-08-005}
{``}\bibinfo{title}{{SUSY searches with dijet events}},{''}
  (\bibinfo{year}{2008}), \bibinfo{note}{{CMS PAS SUS-08-005}}

\bibitem[{PAS(2011{\natexlab{b}})}]{PAS-SUS-11-003}
{``}\bibinfo{title}{{Search for supersymmetry in all-hadronic events with
  $\alpha_{\rm T}$}},{''} (\bibinfo{year}{2011}{\natexlab{b}}),
  \bibinfo{note}{{CMS PAS SUS-11-003}}

\bibitem[{\citenamefont{Randall and Tucker-Smith}(2008)}]{Randall:2008rw}
\bibinfo{author}{\bibfnamefont{L.}~\bibnamefont{Randall}} \bibnamefont{and}
  \bibinfo{author}{\bibfnamefont{D.}~\bibnamefont{Tucker-Smith}},
  {``}\bibinfo{title}{{Dijet Searches for Supersymmetry at the LHC}},{''}
  \bibinfo{journal}{Phys.Rev.Lett.} \textbf{\bibinfo{volume}{101}},
  \bibinfo{pages}{221803} (\bibinfo{year}{2008}), \eprint{0806.1049}

\bibitem[{ATL(2012{\natexlab{a}})}]{ATLAS-CONF-2012-033}
{``}\bibinfo{title}{{Search for squarks and gluinos with the ATLAS detector
  using final states with jets and missing transverse momentum and 4.7 ${\rm
  {\rm fb}^{-1}}$ of $\sqrt{s}$ = 7 TeV proton-proton collision data}},{''}
  (\bibinfo{year}{2012}{\natexlab{a}}), \bibinfo{note}{{ATLAS-CONF-2012-033}}

\bibitem[{ATL(2012{\natexlab{b}})}]{ATLAS-CONF-2012-037}
{``}\bibinfo{title}{{Hunt for new phenomena using large jet multiplicities and
  missing transverse momentum with ATLAS in ${\cal L}$ = 4.7 ${\rm {\rm
  fb}^{-1}}$ of $\sqrt{s}$ = 7 TeV proton-proton collisions}},{''}
  (\bibinfo{year}{2012}{\natexlab{b}}), \bibinfo{note}{{ATLAS-CONF-2012-037}}

\bibitem[{PAS(2012{\natexlab{c}})}]{PAS-SUS-11-020}
{``}\bibinfo{title}{{Search for new physics in events with same-sign dileptons,
  b-tagged jets and missing energy}},{''} (\bibinfo{year}{2012}{\natexlab{c}}),
  \bibinfo{note}{{CMS PAS SUS-11-020}}

\end{thebibliography}

%%%%%%%%%%%%%%%%%%%%%%%%%%%%%%%%%%%%%%%%%%%%%%%%%%%%%%%%%%%%%%%%%%%%%%%%%%%%

\makeatletter
\close@column@grid
\clearpage
\onecolumngrid
\makeatother

%%%%%%%%%%%%%%%%%%%%%%%%%%%%%%%%%%%%%%%%%%%%%%%%%%%%%%%%%%%%%%%%%%%%%%%%%%%%

\appendix

\section*{Appendix: Card File Specification Synopsis}

{ \centering \ovalbox{%
\begin{minipage}{0.975\linewidth}
\vspace{3pt}
{\footnotesize
\begin{Verbatim}[numbers=left,numbersep=6pt]
********* CutLHCO CARD SYNOPSIS Version 2.0 *************************************************************************
          NAME = KEY_1:VALUE_1, KEY_2:VALUE_2; Multiple values go in square brackets; KEY:[VALUE_A,VALUE_B,VALUE_C] *
          Values default to zero-indexed option or UNDEF; Selectors with defined CUTs appear as output statistics   *
********* Object Reconstruction *************************************************************************************
OBJ_ALL # Primary (first read access) classification for all objects; Unclassified objects discarded; (PTM,PRM)     *
OBJ_PHO # Exclusive photon object classification; (PTM,PRM,CUT,JET)                                                 *
OBJ_ELE # Exclusive solo electron object classification; Unclassified objects discarded; (SGN,PTM,PRM,CUT,JET)      *
OBJ_MUO # Exclusive solo muon object classification; Unclassified objects discarded; (SGN,PTC,ETR,PTM,PRM,CUT,JET)  *
OBJ_TAU # Exclusive solo tau object classification; Unclassified objects discarded; (SGN,PTM,PRM,CUT,JET)           *
OBJ_LEP # Zeroth lepton object classification; Default input to latter stages; (EMT,SGN,PTC,ETR,PTM,PRM,CUT)        *
OBJ_JET # Zeroth jet object classification; Default input to latter stages; (HFT,FEM,TRK,MUO,PTM,PRM,CUT)           *
OBJ_LEP_N # Ordered lepton object classification; Index N > 0; (SRC,CMP,EMT,SGN,PTC,ETR,PTM,PRM,CDR,SDR,CUT,ANY)    *
OBJ_JET_N # Ordered jet object classification; Index N > 0; (SRC,CMP,HFT,FEM,TRK,MUO,PTM,PRM,CDR,SDR,CUT,ANY)       *
OBJ_DIL_N # Indexed dilepton object classification; Index N > 0; (LEP,DLS,DLF,DMI,CUT)                              *
    SRC : Source jet/lepton object classifications; [+/- Indices < N]; >=0:Includes; <0:Excludes (Dominant)         *
    CMP : Complementary lepton/jet object classifications; [+/- Indices < N]; >=0:Includes; <0:Excludes (Dominant)  *
    LEP : Local inclusive lepton object classification index; Index N >= 0                                          *
    EMT : Accepted lepton flavors; 0:All; +/-1:Elec; +/-2:Muon; +/-3:Tau; <0:Excludes                               *
    SGN : Accepted lepton sign; 0:All; +1:Positive; -1:Negative                                                     *
    PTC : Accepted muon R=4 cone transverse momentum of adjacent objects range in GeV; [Min,Max]                    *
    ETR : Accepted muon adjacent transverse calorimeter energy & track momentum unitless ratio range; [Min,Max]     *
    HFT : Accepted jet heavy flavor tagging; 0:All; 1:Loose (or Tight); 2:Tight (Only)                              *
    FEM : Accepted jet dimensionless electromagnetic fraction range; [Min,Max]                                      *
    TRK : Accepted jet object composite track count range; [Min,Max]                                                *
    MUO : Accepted jet integration of adjacent poorly isolated muons count range; [Min,Max]                         *
    PTM : Accepted transverse momentum magnitude range in GeV; [Min,Max,Sort]; Sort=1:Ascending; Default:Descending *
    PRM : Accepted pseudo-rapidity modulus range in radians; [Min,Max,Stop]; Stop=1:Terminate upon failure          *
    CDR : Accepted complementary inter-object Delta-R proximity range in radians; [Min,Max]                         *
    SDR : Accepted source intra-object Delta-R proximity range in radians; [Min,Max]                                *
    DLS : Accepted dilepton pair sign classification; -1:Opposite; 0:Any; +1:Same                                   *
    DLF : Accepted dilepton pair flavor classification; 0:Any; 1:Same                                               *
    DMI : Accepted dilepton invariant mass range in GeV; [Min,Max,Sort]; Sort=1:Ascending; Default:Descending       *
    CUT : Event cut if object count is outside given range; [Min,Max,Clip]; Clip=-1:Clip to min; Clip=+1:Pad to max *
    ANY : Events passing any defined current or prior listed jet/lepton classification cut pass all; [Indices < N]  *
    JET : Objects failing exclusive classification are recast as jets; 1:True; Default:False                        *
********* Event Selection *******************************************************************************************
EVT_CAL # Calorimeter based missing transverse energy event specification in GeV; Referenced as MET_N = -1; (CUT)   *
EVT_MET # Zeroth missing transverse energy (MET) event specification in GeV; Includes all objects; (CUT)            *
EVT_MHT # Zeroth scalar transverse momentum sum (MHT) event specification in GeV; Includes all objects; (MAS,CUT)   *
EVT_MEF # Zeroth effective mass (MEF = MET + MHT) event specification in GeV; Includes all objects; (CUT)           *
EVT_MET_N # Indexed MET event specification in GeV; Uses specified leptons & jets; Index N > 0; (LEP,JET,CUT)       *
EVT_MHT_N # Indexed MHT event specification in GeV; Uses specified leptons & jets; Index N > 0; (LEP,JET,MAS,CUT)   *
EVT_MEF_N # Indexed MEF event specification in GeV; Sum of specified MET and MHT; Index N > 0; (MET,MHT,CUT)        *
EVT_RET_N # Indexed unitless ratio (MET_NUM / MET_DEN) of specified factors; Index N > 0; (NUM,DEN,CUT)             *
EVT_RHT_N # Indexed ratio (MET_NUM / sqrt( MHT_DEN )) in GeV^{1/2} of specified factors; Index N > 0; (NUM,DEN,CUT) *
EVT_REF_N # Indexed unitless ratio (MET_NUM / MEF_DEN) of specified factors; Index N > 0; (NUM,DEN,CUT)             *
EVT_DET_N # Indexed absolute vector difference (MET_TWO - MET_ONE) of specified factors; Index N > 0; (ONE,TWO,CUT) *
EVT_LTM_N # Indexed leading lepton & specified missing energy transverse mass in GeV; Index N > 0; (LEP,MET,CUT)    *
EVT_JSM_N # Indexed specified jet s-transverse mass MT2 in GeV; Index N > 0; (JET,CUT)                              *
EVT_JZB_N # Indexed leading dilepton & specified jet Z-balance in GeV; Index N > 0; (JET,DIL,CUT)                   *
EVT_JRM_N # Indexed specified lepton & jet razor mass M_R in GeV; Index N > 0; (LEP,JET,CUT)                        *
EVT_ALR_N # Indexed specified lepton, jet & optional MET unitless alpha_R ratio; Index N > 0; (LEP,JET,MET,CUT)     *
EVT_ALT_N # Indexed specified jet & optional MET, MHT unitless alpha_T ratio; Index N > 0; (JET,MET,MHT,MAS,CUT)    *
EVT_MDP_N # Indexed specified jet & optional MET minimal delta phi angle in radians; Index N > 0; (JET,MET,CUT)     *
EVT_BDP_N # Indexed specified jet & optional MET biased delta phi angle in radians; Index N > 0; (JET,MET,CUT)      *
    LEP : Local inclusive lepton object classification index; Index N >= 0                                          *
    JET : Local inclusive jet object classification index; Index N >= 0                                             *
    DIL : Local inclusive dilepton object classification index; Index N >= 0                                        *
    MET : Missing transverse energy specification index; Index N >= -1; Defaults to local objects when optional     *
    MHT : Scalar transverse momentum sum specification index; Index N >= 0; Defaults to local objects when optional * 
    MEF : Effective mass event specification index; Index N >= 0; Defaults to local objects when optional           * 
    MAS : Masses of individual objects are retained during intermediate computations; 1:True; Default:False         *
    NUM : Numerator content specification index; Refers to one of (MET,MHT,MEF) by context; Index N >= -1 or 0      *
    DEN : Denominator content specification index; Refers to one of (MET,MHT,MEF) by context; Index N >= -1 or 0    *
    ONE : First input content specification index; Refers to one of (MET,MHT,MEF) by context; Index N >= -1 or 0    *
    TWO : Second input content specification index; Refers to one of (MET,MHT,MEF) by context; Index N >= -1 or 0   *
    CUT : Event cut if associated value is outside given range in contextually established units; [Min,Max]         *
*********************************************************************************************************************
\end{Verbatim}
} \vspace{-3pt} \end{minipage}}}

%%%%%%%%%%%%%%%%%%%%%%%%%%%%%%%%%%%%%%%%%%%%%%%%%%%%%%%%%%%%%%%%%%%%%%%%%%%%

\end{document}